\documentclass[aps,prb,twocolumn,superscriptaddress]{revtex4-2}
\usepackage{amssymb, amsmath, bm}
\usepackage{graphicx,onlyamsmath}
\usepackage[utf8]{inputenc}

\usepackage{natbib}
\usepackage{xcolor}
\usepackage[normalem]{ulem}  % ��� ����������� ������
\UseRawInputEncoding

\definecolor{amber-iy}{rgb}{1.0, 0.49, 0.0}
\usepackage{epsfig}
\usepackage{soul}

\usepackage{bm}
\usepackage{graphicx}
\usepackage{amsmath}
\usepackage{amssymb}
\usepackage[utf8]{inputenc}
\usepackage[T1]{fontenc}
\usepackage{color}
\usepackage{upgreek}
\usepackage{subfigure}
\usepackage[unicode=true,colorlinks=true,citecolor=blue]{hyperref}
\usepackage{placeins}
\usepackage{lipsum}
\usepackage{epsfig}
\usepackage[utf8]{inputenc}
\usepackage{wrapfig}
\usepackage{hyperref}

\begin{document}

\title{THz ratchet effect in HgTe interdigitated structures}

\author{I.~Yahniuk}
\affiliation{ Terahertz Center, University of Regensburg, D-93053 Regensburg, Germany}
\affiliation{Center for Terahertz Research and Applications (CENTERA), Institute of High Pressure Physics PAS, Warsaw, 01-142 Poland.}

\author{G.~V.~ Budkin}
\affiliation{Ioffe Institute, 194021 St. Petersburg, Russia}

\author{A.~Kazakov}
\affiliation{International Research Centre MagTop, Institute of Physics, Polish Academy of Sciences, al. Lotnik\'{o}w 32/46, PL02-668, Warsaw, Poland.}

\author{M.~Otteneder}
 \affiliation{ Terahertz Center, University of Regensburg, D-93053 Regensburg, Germany}

 \author{J.~Ziegler}
  \affiliation{ Terahertz Center, University of Regensburg, D-93053 Regensburg, Germany}

 \author{D.~Weiss}
 \affiliation{ Terahertz Center, University of Regensburg, D-93053 Regensburg, Germany}

\author{N.~N.~Mikhailov}
\affiliation{Institute of Semiconductor Physics, Siberian Branch, Russian Academy of Sciences, pr. Akademika Lavrent'eva 13, Novosibirsk, 630090 Russia.}

\author{S.~A.~Dvoretskii}
\affiliation{Institute of Semiconductor Physics, Siberian Branch, Russian Academy of Sciences, pr. Akademika Lavrent'eva 13, Novosibirsk, 630090 Russia.}

\author{T.~Wojciechowski}
\affiliation{International Research Centre MagTop, Institute of Physics, Polish Academy of Sciences, al. Lotnik\'{o}w 32/46, PL02-668, Warsaw, Poland.}

\author{V.~V.~Bel'kov}
\affiliation{Ioffe Institute, 194021 St. Petersburg, Russia}

\author{W.~Knap}
\affiliation{Center for Terahertz Research and Applications (CENTERA), Institute of High Pressure Physics PAS, Warsaw, 01-142 Poland.}
\affiliation{Laboratoire Charles Coulomb, Universit\'{e} de Montpellier, Centre National de la Recherche Scientifique, 34095 Montpellier, France.}

\author{S.~D.~Ganichev}
\affiliation{ Terahertz Center, University of Regensburg, D-93053 Regensburg, Germany}
\affiliation{Center for Terahertz Research and Applications (CENTERA), Institute of High Pressure Physics PAS, Warsaw, 01-142 Poland.}

\date{\today}

\begin{abstract}
The emergence of ratchet effects in two-dimensional materials is strongly correlated with the introduction of asymmetry into the system. In general, dual-grating-gate structures forming  lateral asymmetric superlattices provide a suitable platform  for studying this phenomenon.  Here, we report on the observation of ratchet effects in HgTe-based dual-grating-gate structures  hosting  different band structure properties.  Applying polarized terahertz laser radiation we detected    linear and polarization independent ratchets, as well as an radiation-helicity driven circular ratchet effect. Studying the ratchet effect in devices made of quantum wells (QWs) of different thickness we observed that the magnitude of the signal substantially increases with decreasing QW width with a maximum value for devices made of QWs of critical thickness hosting Dirac fermions. Furthermore, sweeping the gate voltage amplitude we observed sign-alternating oscillations for  gate voltages corresponding to $p$-type conductivity. The amplitude of the oscillations is more than two orders of magnitude larger than the signal for $n$-type conducting QWs. The oscillations and the signal enhancement are shown to be caused by the complex valence band structure of HgTe-based QWs. These peculiar features of the ratchet currents make these materials an ideal platform for the development of THz applications.

\end{abstract}

%\pacs{\textbf{Check:} 73.21.Fg, 73.43.Lp, 73.61.Ey, 75.30.Ds, 75.70.Tj, 76.60.-k} 
                            
%\keywordsterahertz ratchet, HgTe lateral superlattices}

\maketitle

\section{\label{sec:Int}Introduction}
Mercury telluride based heterostructures belong to the most widely used materials  for sensitive and fast infrared/terahertz 	(IR/THz) 	detectors~\cite{Capper1997,Norton2002,Henini2002,Rogalski2005,Dvoretsky2010,Downs2013,Rogalski2018,Vanamala2019} 	and are among the most promising materials to realize high quality 	topological insulators (TIs)~\cite{Moore2010,Hasan2010,Zhang2011}. The reason	for that is the wide tunability of the energy spectrum of these	materials including the possibility of realizing an inverted band 	structure in HgTe. The latter is a crucial condition for the formation of helical edge and surface states~\cite{Moore2010,Hasan2010,Zhang2011}. This is also supported by the well developed technological 	processes originally motivated by the fabrication of detectors, which has been adapted for the growth of high-quality TI materials.  This includes the  possibility to obtain high carrier mobility  while at the same time contributions  from three-dimensional carriers in the bulk can be largely suppressed~\cite{Becker2003,Dvoretsky2010,Koenig2007,Roth2009,Dantscher2015,Dantscher2017}. Thus HgTe-systems allow one combining the excellent performance of HgTe-based IR/THz sensors and advantages of topological systems, in particular, obtaining photon helicity sensitive photoresponses~\cite{Wittmann2010,Dantscher2017}. In the last decade, it has been demonstrated that ratchet effects in two-dimensional electron systems with lateral superlattices can be used for efficient detection of THz radiation and may even provide new functionalities, such as all-electric detection of the radiation Stokes parameters operating up to room temperature~\cite{Danilov2009}.  The ratchet effect, demonstrating strong photoresponse and considered as a candidate for efficient detection of THz radiation, has been observed and	investigated in various 2D semiconductor structures with parabolic energy dispersion~\cite{Olbrich2009,Olbrich2011,Ivchenko2011,Popov2011,Kannan2011,Otsuji2013,Boubanga2014,Faltermeier2015,Rupper2018,Yu2018,Notario2020,Sai2021} and in monolayer graphene~\cite{Olbrich2016}, but so far has not been detected  in HgTe-based systems. Such a study, however,  allows to combine advantages of the ratchet effect and the superior properties of HgTe-based materials for the infrared/terahertz radiation detection, as well as exploring the physical properties of HgTe-based QWs. 

Here we report on the observation and study of  polarization-dependent ratchet effects in HgTe-based quantum wells of  different thicknesses. The effect was studied in QWs with superimposed dual-grating gate (DGG), lateral asymmetric superlattices excited by normally incident terahertz laser radiation. The magnitude and direction of the ratchet current are shown to be dependent both on the polarization state of the radiation and on the lateral asymmetry determined by the gate voltages applied to the two subgates. By varying the radiation polarization we observed that the THz ratchet effect has three current contributions: polarization-insensitive, linear-, and circular- ratchet ones. While the second one can be excited by  linearly polarized radiation and is sensitive to the relative orientation of the electric field vector and the source-drain direction, the third one is driven by the radiation helicity and has opposite signs for right- and left-handed circularly polarized radiation. Measurements of the ratchet currents in HgTe QWs of different thicknesses allowed us to study ratchet effects in HgTe-based  QWs featuring different band structure properties. Notably, the highest helicity-driven ratchet current was detected in a system with Dirac fermions. Studying the gate voltage dependence of the ratchet effect we, unexpectedly, observed emergence of high sign alternating oscillations of the photoresponse at high negative gate voltages corresponding to $p$-type conductance. Our theoretical analysis demonstrated that the oscillations are caused by  shifting   the Fermi energy across the well-separated multiple valence subbands, which in turn results in  oscillations of the density of states. The developed theoretical model, which takes into account the Seebeck ratchet mechanism and considers tiny details of the energy dispersion, describes the experimental results well.

  \begin{table*}[t]
	\centering
	\begin{tabular}{ccccccccc}
		\hline
		Devices & $d_{QW}$ (nm) & Barrier composition &  $L\times W$ ($\mu m^2$) &  $d_1$ ($\mu m$) & $d_2$ ($\mu m$) & $a_1$ ($\mu m$) & $a_2$ ($\mu m$) & $d$ ($\mu m$)\\
		\hline
		\hline
		D1 &8.0 & Hg$_{0.23}$Cd$_{0.77}$Te & 72$\times$31 & 0.75 & 2.25 & 1.0 & 3.5 & 7.5\\		
		D2 &7.0 & Hg$_{0.28}$Cd$_{0.72}$Te & 85$\times$20 & 0.5 & 1.5 & 0.5 & 2.5 & 5.0\\	
		D3 &6.3 &  Hg$_{0.42}$Cd$_{0.58}$Te & 50$\times$19 & 0.5 & 1.5 & 0.5 & 2.5 & 5.0\\
		D4 &7.0 &  Hg$_{0.28}$Cd$_{0.72}$Te & 75$\times$15 & 1.25 & 3.0 & 0.75 & 3.0 & 8.0\\
		D5 &8.0 &  Hg$_{0.23}$Cd$_{0.77}$Te & 75$\times$15 & 1.25 & 3.0 & 0.75 & 3.0 & 8.0\\
		D6 &8.0 &  Hg$_{0.23}$Cd$_{0.77}$Te & 50$\times$7 & 0.75 & 2.0 & 0.5 & 2.0 & 5.25\\
		D7 &7.0 &  Hg$_{0.28}$Cd$_{0.72}$Te & 72$\times$31 & 0.75 & 2.25 & 1.0 & 3.5 & 7.5\\
		\hline
	\end{tabular}
	\caption{\label{tab:1} Basic parameters of the  structures investigated.}
\end{table*}
  
  	\begin{figure*}[ht]  %FIG. 1
	\includegraphics [width=0.65\linewidth, keepaspectratio] {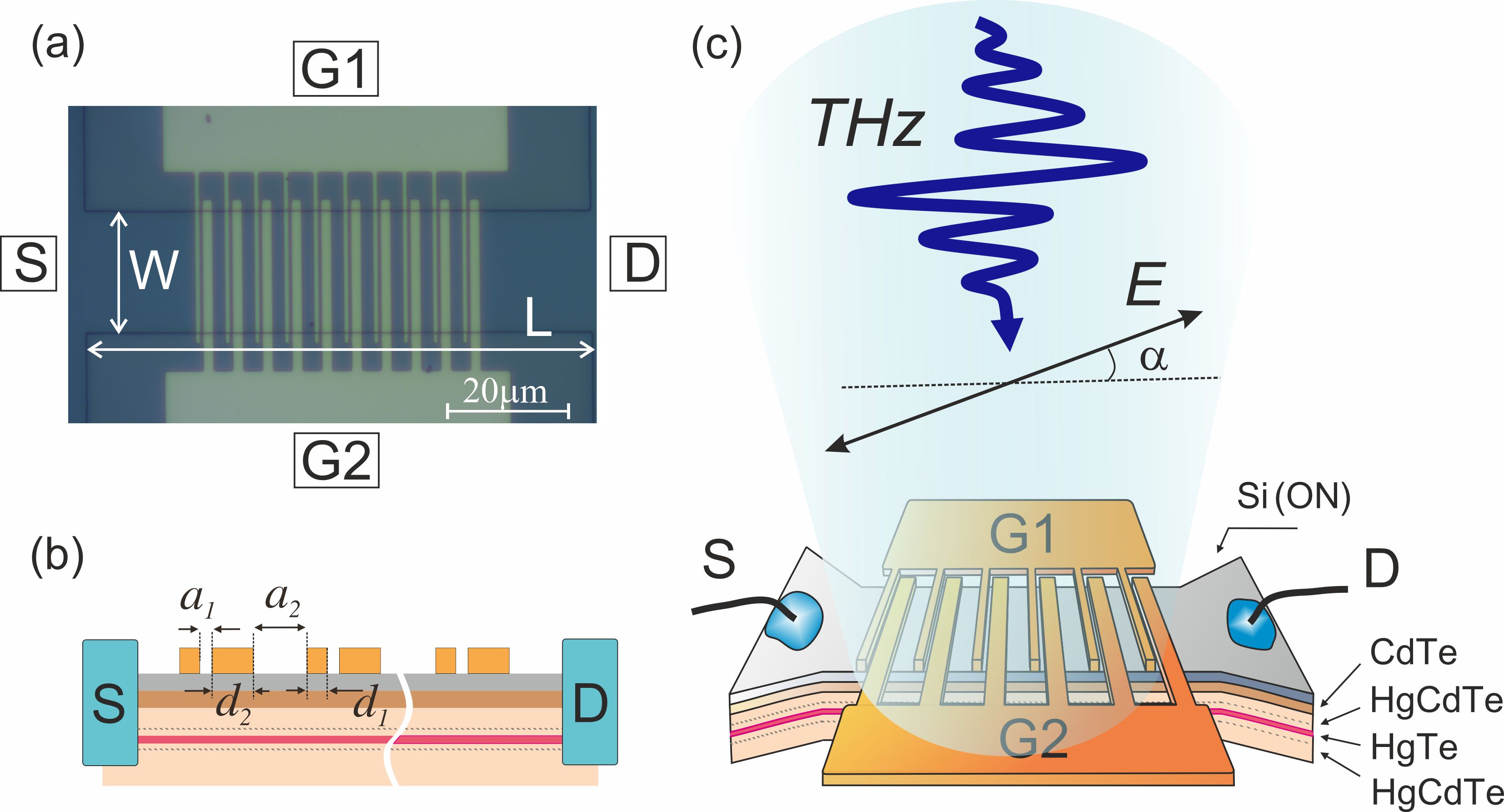} 
	\caption{\label{Fig:1} (a)~Picture of the interdigitated gate electrodes deposited on the  HgTe/HgCdTe heterostructure. All devices have four terminals: source(S), two gates(G1, G2), and drain (D). (b)~A cross-section  of the asymmetric finger gate structure consisting of stripes of widths $d_1$ and $d_2$ and spacing $a_1$ and $a_2$  forming a superlattice with periodicity $d = d_1 + a_1 + d_2 + a_2$.   (c)~Sketch of the HgTe/HgCdTe heterostructure under incident THz radiation indicating also the QW layer sequence.}
\end{figure*}

\subsection{Experimental technique}

Our devices are made   HgTe/CdHgTe quantum well (QW) structures grown by molecular beam epitaxy on (013)-oriented GaAs substrates. The heterostucture parameters, such as QW thickness ($d_{QW}$) and barrier composition are presented in Table~\ref{tab:1}. The asymmetric inter-digitated DGG structures were fabricated by electron beam lithography (EBL). Wet Br-based etching   is used to define the  channel geometries, having a channel length ranging from 50 to 75~$\mu$m and a channel width  between 7 and 31~$\mu$m. Plasma-enhanced chemical vapor deposition was used to deposit  140~nm \rm{Si(ON)} insulation, separating the HgTe/CdHgTe heterostructure and Ti/Au finger gates. The full description of the growth, characterization and device preparation can be found in Ref.~\cite{Majewicz2014,Majewicz2019}.  Figures~\ref{Fig:1}(a) and~\ref{Fig:1}(b) show an optical image of one of the DGG structures and a schematic illustration of the cross-section of the device. The grating consists of  periodically repeating asymmetric supercells   which are separated by    spacings $a_1$ and $a_2$. The asymmetry of the DGG structure stemming  from the inequallity of stripe widths ($d_1 < d_2$) and stripe separations ($a_1 < a_2$)  is crucial for the generation of the ratchet effect. Below we denote the direction perpendicular to the metal fingers as $x$-direction.

All thin stripes are connected by a metal film forming the gate G1, and all interconnected wide stripes form the gate G2. This allows us to create an asymmetric periodic electrostatic potential acting on 2DEG by applying different voltages to the subgates G1 ($V_{\rm G1}$) and G2 ($V_{\rm G2}$).  Table~I. presents the geometrical parameters of the DGG and period of the supperlattice, $d = d_1 + a_1 +d_2 + a_2$. Ohmic contacts to source, drain, and gates were fabricated by In soldering. To characterize the structures we performed transport and magnetotransport measurements. To measure the electrical resistance $R_{SD}$ we have used SR830 lock-in amplifiers with low modulation frequency (13~Hz) and $0.1~\upmu$A current amplitude. All transport studies were carried out at $T=4.2$~K. The source-drain resistance as a function of the gate voltage exhibits a clear maximum at negative values of the gate voltages $V_{\rm G1}$ or $V_{\rm G2}$.  This demonstrates that besides the controllable modification of the lateral asymmetry, sweeping  the gate voltages allows us to change the type of conductivity beneath the gates. The variation of the carrier density of the 2DEG in the HgTe QWs from $n$- to $p$-type occurs for a Fermi energy position in the insulating band gap. In this case, the   resistance $R_{SD}$ shows a maximum which corresponds to a change of the sign of the Hall coefficient and identifies the charge neutrality point (CNP). For the investigated structures with 6.3, 7.0, and 8.0~nm QW thickness at zero gate voltage, the carrier concentrations are in the range of $(1.3 \div 7.4) \times 10^{11}$cm$^{-2}$, and mobilities are about $ (5.0 \div 8.5) \times 10^4$~cm$^2$/Vs.

To excite the ratchet effects we applied normal-incident terahertz radiation of a continuous wave (cw)  optically pumped molecular laser,  see Fig. \ref{Fig:1}(c).  The laser operated at frequency $f= 2.54$\,THz (wavelength $\lambda = 118~\upmu$m, photon energy $\hbar\omega = 10.5$~meV). The samples were placed in an optical cryostat with $z$-cut quartz and TPX (4-methyl-1-pentene) windows.  In order to block visible and near-infrared radiation, the windows were additionally covered with a black polyethylene film. The laser beam was focused using off-axis parabolic mirrors. The radiation  power at the sample position $P$ was about 20~mW and was monitored during the measurements by a pyroelectric detector.  The beam had an almost Gaussian shape with  a spot diameter of around 1.5~mm,  which is much larger than the area of  the DGG structure. The spatial distribution of incoming THz beam was controlled by a pyroelectric camera. More details on the system can be found in Refs.~\cite{Dantscher2015,Dantscher2017,Shalygin2007,Plank2016}. The radiation was modulated by a chopper with a frequency of 35~Hz and  standard lock-in technique was used to detect the photoresponse,  measured as  voltage drop $V_{ph}$ across the sample resistor $R_S$. All measurements were performed at helium temperature $T = 4.2$~K.

To explore the polarization dependence of the THz radiation induced signal we controllably varied the radiation Stokes parameters. The polarization was modified by placing a crystal quartz $\lambda/4$-plate in front of the sample, which was rotated by a phase angle $\varphi$ between the $c-$axis of the plate and the  electric field vector of the laser radiation. This allowed us to vary the degree of circular polarization $P_{\rm circ} = \sin 2\varphi$, which corresponds to the Stokes parameter $s_3$,  as well as the degrees of the linear polarization (these Stokes parameters are $s_1/s_0 = \sin 4\varphi /2$ and $s_2/s_0 = (1+\cos 4\varphi)/2$)~\cite{Belkov2005}). Additionally, we performed measurements in which the vector of the electric field $\bm E$  was rotated by a $\lambda$/2-plate. In this case, the azimuth angle $\alpha$ between $\bm E$ and $x$-direction determines the orientation of  the incident radiation regarding the gate stripes,  see Fig. \ref{Fig:1}(c), and the Stokes parameters are described by $s_1/s_0 = \sin 2\alpha$ and $s_2/s_0 = \cos 2\alpha$~\cite{Belkov2005}.

\begin{figure}[h]  %FIG. 2
	\includegraphics [width=\columnwidth, keepaspectratio] {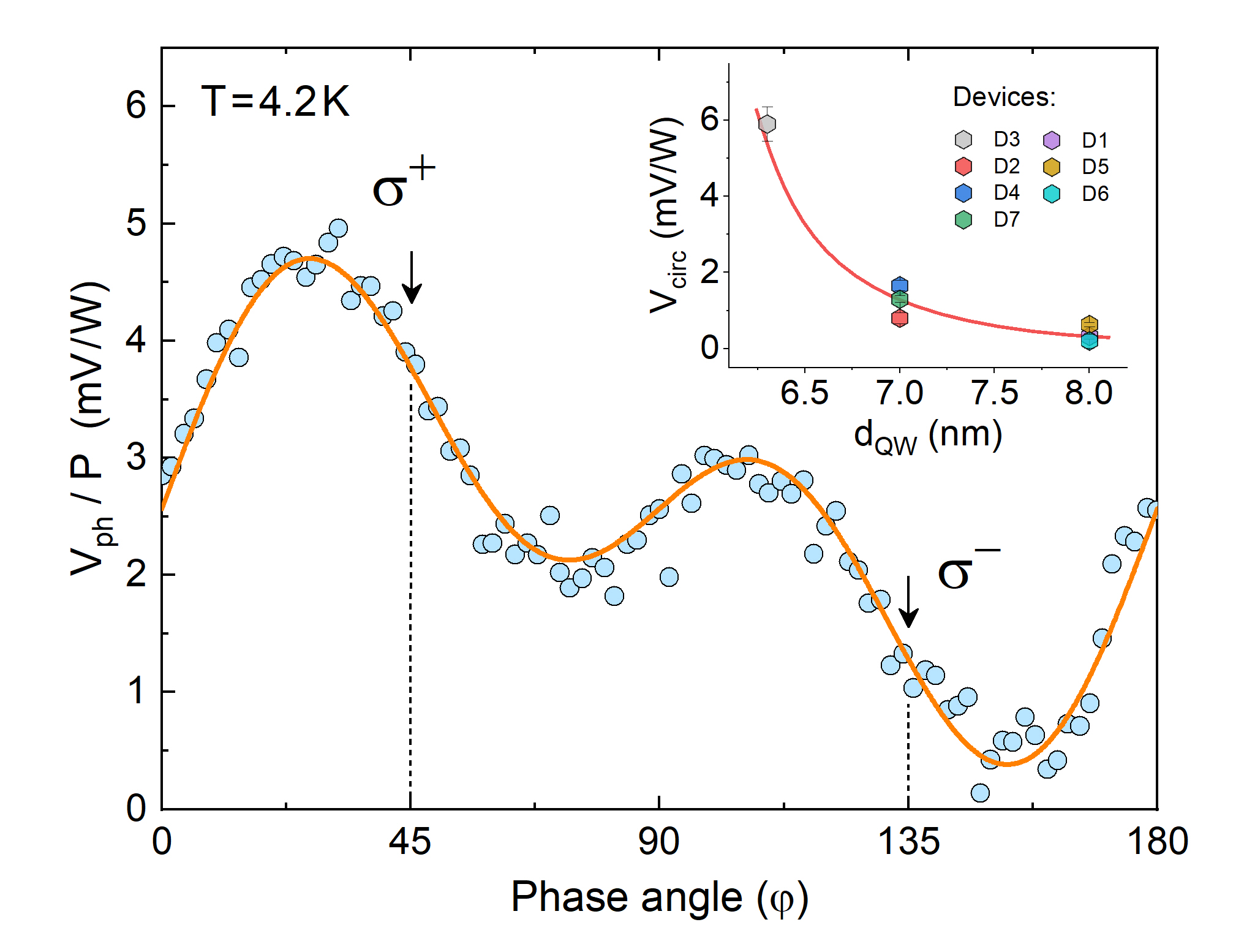} 
	\caption{\label{Fig:4} Helicity dependence of the photovoltage of sample D7 normalized to power  under normal incident radiation  with frequency $f=2.54$~THz.  Measurements are carried out at $T=4.2$~K and zero gate voltages. Arrows correspond to right-handed ($\sigma^+$) and left-handed~($\sigma^-$) circular polarizations. The orange line is a fit  according to Eq.~\eqref{Helicity}.  The inset indicates the circular photoresponse  as a function of the QW thickness for different devices (D1~-~D7).}
\end{figure}

\begin{figure}[h]  %FIG. 3
	\includegraphics [width=1.0\columnwidth, keepaspectratio] {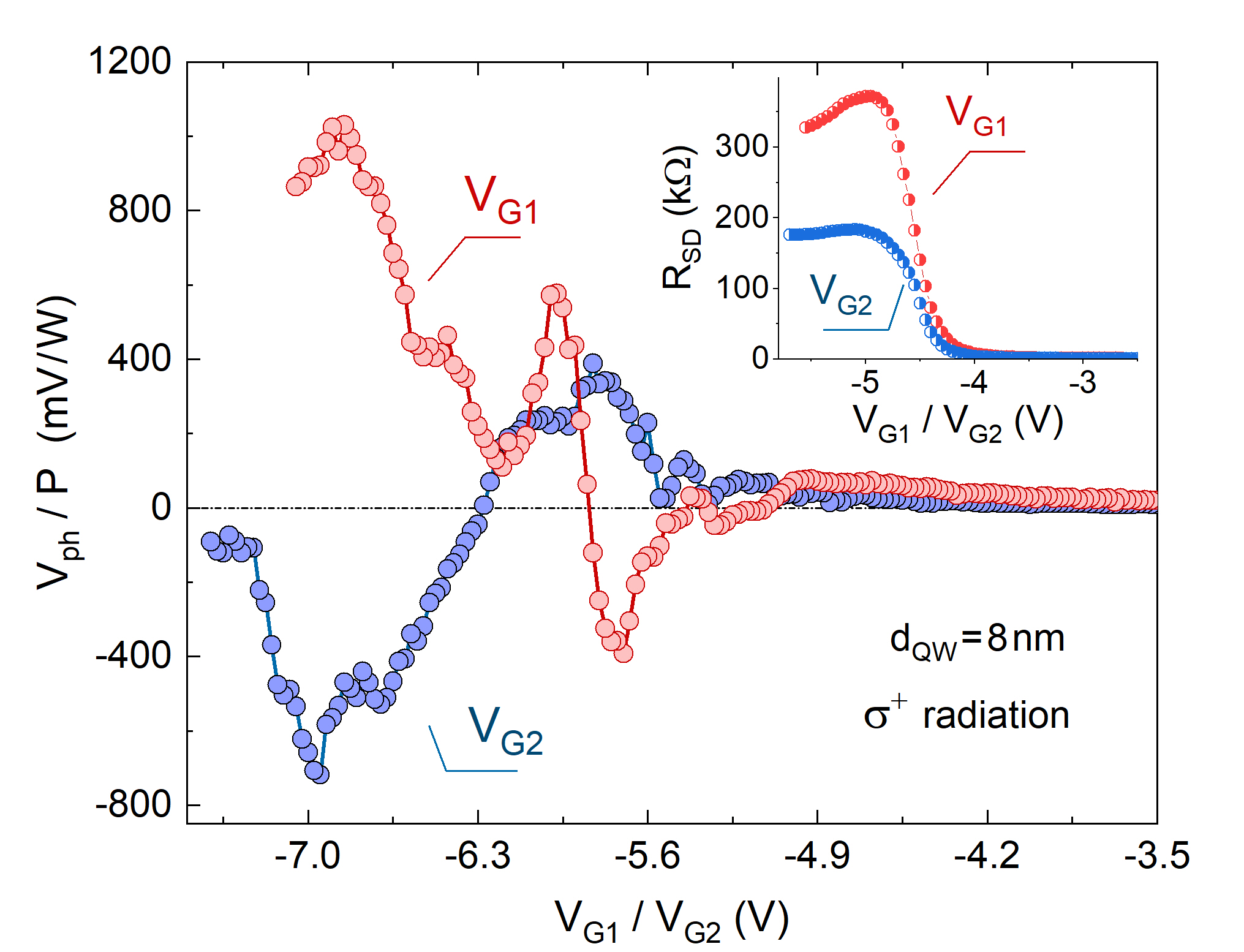}
	\caption{\label{Fig:8}   Photovoltage normalized to power as a function of the gate voltage for device~D6.  Red and blue lines correspond to gate voltages applied to electrodes with thin ($G1$) and thick ($G2$) fingers, respectively. Measurements are performed under normal incident right-handed circularly polarized radiation ($\sigma^+$) with frequency  $f=2.54$~THz   and $T=4.2$~K. The inset shows the  resistance  $R_{SD}$ vs   gate voltage  applied to thin stripes $V_{\rm G1}$ ($V_{\rm G2}=0$, red symbols) and to thick stripes  $V_{\rm G2}$ ($V_{\rm G1}=0$,~blue symbols).}
\end{figure}

\begin{figure}[h]   %FIG. 4
	\includegraphics [width=\columnwidth, keepaspectratio] {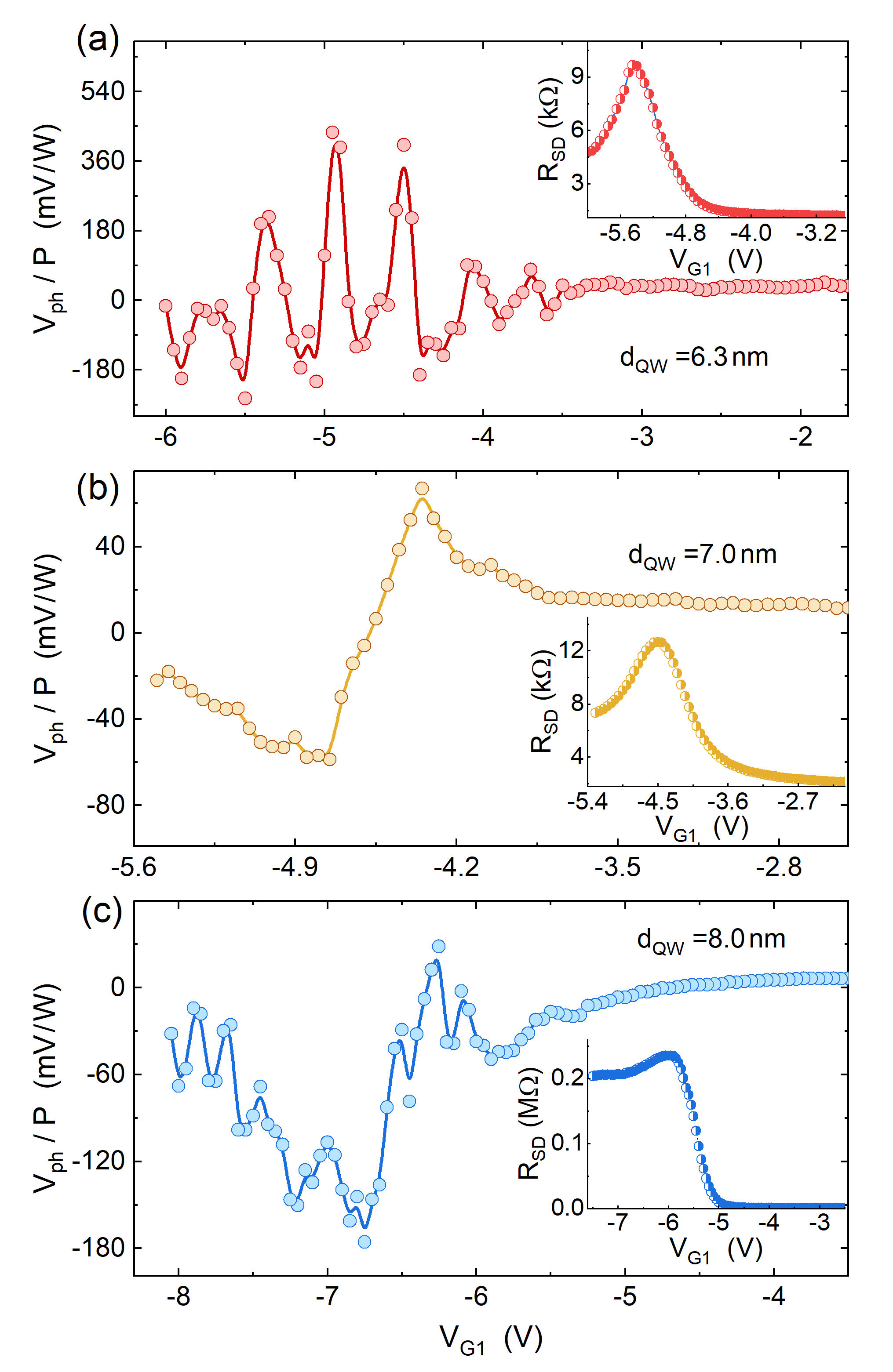} 
	\caption{\label{Fig:6} (a-c)~Normalized photovoltages in devices D3, D2, and D5  versus gate voltage $V_{\rm G1}$. All data are obtained at normal incident of linearly polarized radiation   frequency $f = 2.54$~THz at $\alpha_{max}$, and $T = 4.2$~K. Here $\alpha_{max}$ is the azimuth angle at which the signal achieves its maximum. The solid lines are smoothed curves. Insets show results of transport measurements in HgTe QW devices D3, D2, and D5. The resistances were measured between source and drain as a function of the gate voltage  $V_{\rm G1}$.
}
\end{figure}

\subsection{Results}
 
Figure~\ref{Fig:4} shows the dependence of the photosignal excited in sample D7 (QW thickness $d_{QW}=7$~nm) on the phase angle $\varphi$. The signal change $V_{ph}$ when changing the polarization  is characteristic for the ratchet effects and can be well described by the function~\cite{Olbrich2016,Ivchenko2011}
\begin{equation}
\label{Helicity}
V_{ph}(\varphi) = V_0 + V_\mathrm{L1}\frac{\sin 4\varphi}{2} + V_\mathrm{L2}\frac{1 + \cos 4\varphi}{2} + V_\mathrm{circ} \sin 2\varphi  \,\,,
\end{equation}
where the fitting parameters $V_0, V_\mathrm{L1}$, $V_\mathrm{L2}$, and  $V_\mathrm{circ}$ describe the polarization independent ratchet contribution ($V_0$, also called Seebeck ratchet),  the linear ratchet effect~($V_\mathrm{L1}, V_\mathrm{L2}$), and the circular ratchet effect~($V_\mathrm{circ}$), respectively. This characteristic polarization dependence has been detected for all samples and  give a first hint that we have to do here with a ratchet effect. Note that the above equation reflects the linear combination of the radiation Stokes parameters $s_0, s_1, s_2$,  and $s_3$ with different weights. For the right- ($\sigma^+ $) and left-($\sigma^- $) handed circularly polarized radiation the Stokes parameter $s_3$ changes its sign, whereas $s_1$ and $s_2$ vanish and the first polarization-independent term in Eq.~(\ref{Helicity}) remains constant. Consequently, the circular photoresponse can be calculated as the odd part of  the voltage signal with respect to the  radiation helicity
\begin{equation}
\label{circ}
V_\mathrm{circ} = \frac{V^\mathrm{\sigma^+} - V^{\sigma^-}}{2} \,\,,
\end{equation}
where $V^\mathrm{\sigma^+}$ is   the measured photosignal at  right-handed circular polarization ($\varphi =45^{\circ}$), and $V^\mathrm{\sigma^-}$ corresponds to the measured photosignal at left-handed circular polarization ($\varphi =135^{\circ}$)
The inset in Fig.~\ref{Fig:4} compares the circular contributions for the studied devices having different quantum well thickness. One can see that the most pronounced feature of the ratchet effect is observed in device D3  ($d_{QW}=6.3$~nm) with a quantum well thickness close to the critical one, i.e. hosting two-dimensional massless fermions. 

For linearly polarized radiation the last contribution in Eq.~(\ref{Helicity}) vanishes and the polarization dependences of the second and third terms are given by $s_1/s_0 = \sin 2\alpha$ and $s_2/s_0 = \cos 2\alpha$ with  amplitudes  $V_\mathrm{L1}$, and $V_\mathrm{L2}$, respectively (not shown). 
To ensure that the detected signal is caused by the ratchet effects we studied the photoresponse as a function of the top gate voltages. According to theory~\cite{Ivchenko2011}   the ratchet contributions should reverse sign upon inversion of the lateral asymmetry parameter given by~\cite{Ivchenko2011}
\begin{equation}
\label{Xi}
\Xi = \overline{|E(x)|^2\frac{dV(x)}{dx}\, .}
\end{equation}
Here the overline means an average over the coordinate perpendicular to the DGG stripes. $V(x)$ is the electrostatic potential induced by lateral superlattice, $E(x)$ is the spatially modulated near electric field of radiation acting on charge carriers in QW. Consequently, it is controlled by the potential variation $dV(x)/dx$, which is determined by the voltages applied to the individual gates, $V_{\rm G1}$ and $V_{\rm G2}$. In order to tune the lateral asymmetry, we hold one of the gates at zero bias and vary the gate voltage on the other. Figure~\ref{Fig:8} demonstrates that the photosignals obtained upon variation of $V_{\rm G1}$ and $V_{\rm G2}$, i.e. by the interchange of gate voltage polarities at narrow and wide gates, have consistently opposite signs. This observation, exemplary presented for the device D6 and left-handed circularly polarized radiation, agrees well with the signature of ratchet photovoltages, $V_{ph} \propto \Xi$, and proves that ratchet effects are responsible for it.

Strikingly, for   high negative potentials applied to the top gates ($V_{\rm G1}$ or $V_{\rm G2}$),  the photosignal exhibits sign-alternating oscillations when varying  the top gate voltage. The number of the oscillations in the ratchet photosignal depends on the QW thickness, see Fig.~\ref{Fig:6}. The photosignal oscillations are   detected for both linear- and circular- ratchet effects. Note that the sample resistance varies smoothly with  the gate voltage and only exhibits a peak at the CNP corresponding to the transition from $n$- to $p$-type conductivity, see insets in Fig.~\ref{Fig:6}. We also emphasize that the oscillations appear for   gate voltages corresponding to the $p$-type conductivity. 

The overall features of the observed THz photoresponse, in particular its proportionality to the lateral asymmetry parameter (see Fig.~\ref{Fig:8}), clearly indicates that it is caused by the ratchet effect. Measurements on QWs with different thickness reveal that the magnitude of the ratchet effect in HgTe-based structures increases substantially (more than   an order of magnitude)   with decreasing and becomes maximal for the QW with critical thickness, see the inset in Fig.~\ref{Fig:4}. This we ascribe to the qualitative change of the energy dispersion, which changes from  parabolic with inverted band structure to a linear one for the critical thickness (6.4~nm). 

The key result of the present work is the observation  of the unexpected ratchet current  oscillations when varying  the top gate voltages, see Figs.~\ref{Fig:8} and~\ref{Fig:6}. While the sample resistance shows the conventional behavior for   narrow band HgTe systemswith a resistance  maximum at $V_{CNP}$  (see insets in Figs,~\ref{Fig:8} and~\ref{Fig:6}),   the ratchet responses in all samples exhibit  sign-alternating oscillations for high negative top gate voltages . Furthermore, the amplitude of the ratchet effects drastically increases by more than two orders of magnitudes, as compared to that detected for lower gate voltages at which the samples have  $n-$type conductivity, see Figs.~\ref{Fig:8} and~\ref{Fig:6}.  While oscillations and increase of the ratchet current magnitude are detected in all samples, the number of oscillations is specific for each sample and is maximum for the quantum well structures with critical $d_{QW}$ in which the energy gap is very small or vanishing, see Fig.\ref{Fig:6}(a). 
Whereas  giant oscillations   have been detected in  DGG structures in an external magnetic field~\cite{Sai2021,Faltermeier2017,Budkin2016,Faltermeier2018,Hubmann2020}, no oscillations in the ratchet voltage have been observed at zero magnetic field. In the next section, we discuss the origin of these photocurrent oscillations in HgTe-based DGG structures, thus revealing peculiar features of the  valence band energy spectrum in HgTe quantum wells with different QW thicknesses. 

\section{DISCUSSION}
 
To explore the origin of the observed oscillations, we develop a theory for ratchet currents in the HgTe quantum wells with $p$-type conductivity. To be specific, we  consider the polarization-independent contribution of the photocurrent, which is generated via  energy relaxation of the radiation-heated holes (the Seebeck ratchet). A detailed description of this mechanism can be found in Refs.~\cite{Ivchenko2011,Faltermeier2017}. The theory of polarization-dependent ratchet currents is out of the scope of the present manuscript. However, it can be developed using kinetic theory from~\cite{Durnev2021,Nalitov2012} taking into account the complicated band structure of the QW.

\begin{figure*}[ht]   %FIG 5
\begin{center}
\includegraphics[width=0.7\linewidth]{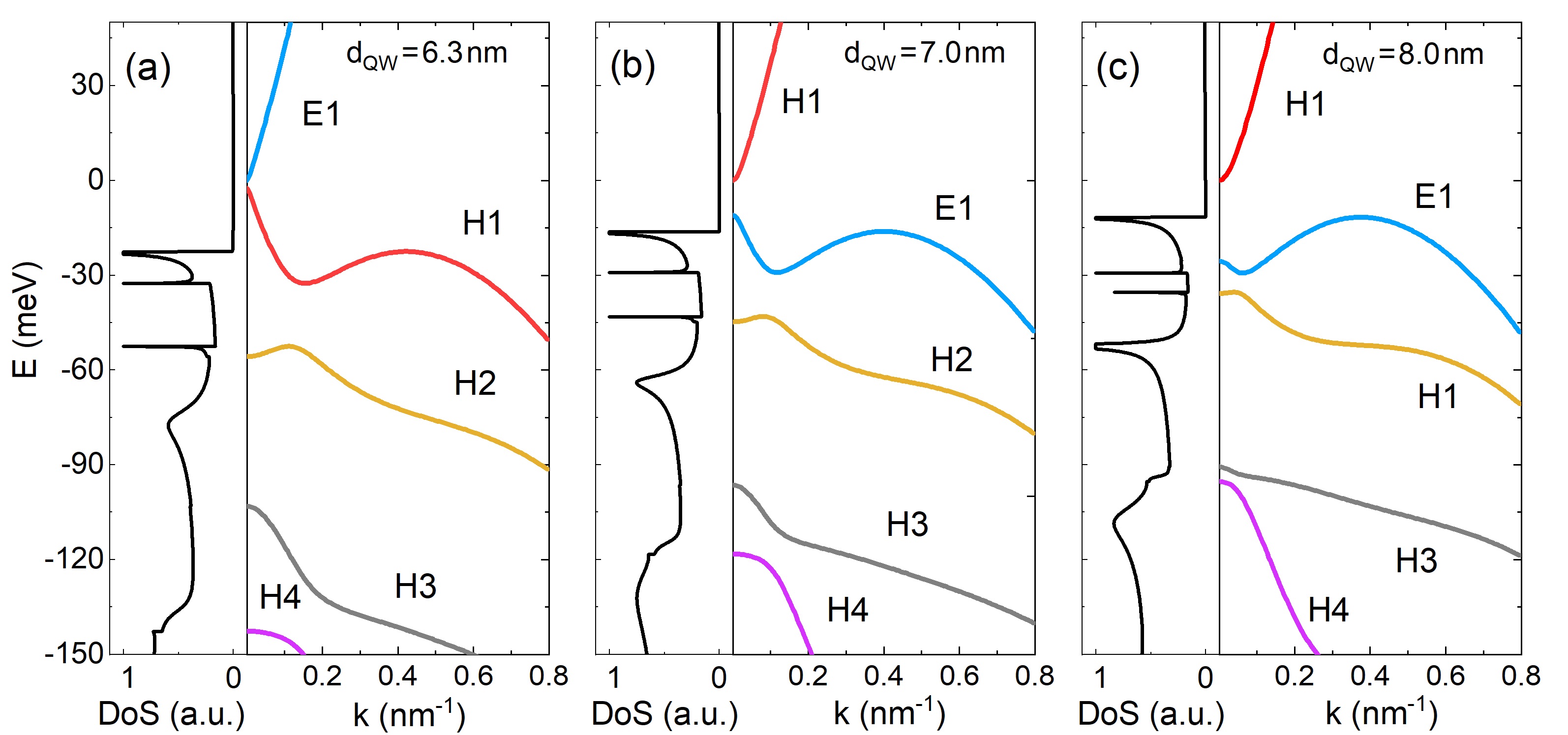}
\end{center}
\caption{The calculated energy spectra of the quantum wells of width $6.3$~nm (panel a),  $7$~nm (panel b), and $8$~nm (panel c) shown by colored lines. Black lines on the left side of the panels show the DoS for the corresponding energy spectra obtained in isotropic approximation.}
\label{fig_spectrum}
\end{figure*}

The knowledge of the band structure is crucially needed for the theory of electric current generation and understanding of the oscillation origin. The energy dispersion of differently wide quantum wells, shown in Fig.~\ref{fig_spectrum}, are calculated within the framework of an 8-band $\bm{k} \cdot \bm{p}$~model described in detail in Ref.~\cite{Dantscher2015}. The energy dispersion of the hole subbands varies greatly with the change of quantum wells width. The  $6.3$~nm quantum well has bandgap close to zero, for $7$~nm the band gap between hole and electrons is increased, at $8$~nm bottom electron and top hole subbands are pushed even further away at the $\Gamma$-point, however, the quantum well has an indirect bandgap due to the anticrossing between first and second hole subbands. We consider the case of low temperatures, when free charge carriers are described by a nearly degenerate distribution function. In this case, the electronic properties are determined by carriers near the Fermi energy. For certain values, the Fermi level crosses the spectrum at several  wave vector values, so that there are essentially multiple types of carriers, each with its own Fermi velocity and density, all of which contribute to both conductivity and ratchet current. Due to the anticrossing in the valence band multiple Van Hove singularities are observed in the density of states (DoS). This is shown in the left-hand sides of Figs.~\ref{fig_spectrum}(a) --~\ref{fig_spectrum}(c). Both momentum relaxation time and charge carrier density change significantly when the Fermi energy is close to these singularities. As will be shown below, these are the reasons for the drastic change of the ratchet current magnitude and direction.

Now we turn to the microscopic theory of the Seebeck ratchet current, which is based on models previously developed in Refs.~\cite{Budkin2016,Olbrich2016}.  The Seebeck ratchet current emerges when there is a phase difference between the  in-plane component of the static electric field $\partial V(x)/\partial x$ caused by the electric potential induced by DGG and the spatially modulated  free charge carrier  heating caused by the electric near field radiation. To calculate the ratchet current first we split the  time-independent part of  the distribution function as ${f_{\bm{p},\nu}=f^+_{\bm{p},\nu}+f^-_{\bm{p},\nu}}$, where $\bm{p}$ is the charge carrier momentum, $\nu$ is the subband index, and $f^+_{\bm{p},\nu}$ and $f^-_{\bm{p},\nu}$  are even and odd on the momentum parts of the distribution
function, respectively. The Boltzmann kinetic equation leads to the following relation 
\begin{equation}
-\dfrac{f^-_{\bm{p},\nu}}{\tau_p}=v_x\dfrac{\partial f^+_{\bm{p},\nu}}{\partial x}- \dfrac{\partial V}{\partial x} \dfrac{\partial f^+_{\bm{p},\nu}}{\partial p_x}\:,
\end{equation}
where $\tau_p$ is the momentum relaxation time, $\bm{v}$ is  the velocity defined as $\partial \varepsilon_{\bm{p},\nu}/\partial \bm{p}$, and $\varepsilon_{\bm{p},\nu}$ is the charge carrier energy. The Seebeck ratchet current is obtained by   summing over all   momentum and subband indices $ j_x=2e \sum_{\bm{p},\nu} v_x f^-_{\bm{p},\nu}$, where the factor $2$ stands for spin degeneracy. The ratchet current linear in $V(x)$ and radiation intensity  is given by
\begin{equation}
j_x=-\dfrac{1}{e} \sigma \dfrac{\partial V}{\partial x}-\dfrac{1}{e} \dfrac{\partial}{\partial x} S \:,
\end{equation}
where conductivity $\sigma$ and $S$ are given by
\begin{equation}
\label{conducivity_diff}
\sigma=e^2 \sum_{\bm{p},\nu} v^2 \tau_p \left( -\dfrac{\partial f^+_{\bm{p},\nu}}{\partial \varepsilon_{\bm{p},\nu}}\right)\:,
\quad\quad S=e^2 \sum_{\bm{p},\nu} v^2 \tau_p  f^+_{\bm{p},\nu}\:.
\end{equation}
Due to fast electron-electron interaction the electron gas is locally thermalized and $f^+_{\bm{p}}$ is described by the Fermi-Dirac distribution function
\begin{equation}
f^+_{\bm{p},\nu }=\left[\exp\left(\dfrac{\varepsilon_{\bm{p},\nu}+V(x)-\varepsilon_F-\delta \varepsilon_F(x)}{T+\delta T(x)}\right)+1\right]^{-1}\:,
\end{equation}
where $\delta T(x)$ and $\delta \varepsilon_F(x)$ are small nonequilibrium corrections of the temperature and Fermi energy induced by radiation. Since no current flows without radiation or at $V(x)=0$, these conditions yield
\begin{equation}
\label{zero_current}
\sigma_0=\dfrac{\partial S_0}{\partial \varepsilon_F}\:,\quad\quad 0=
\dfrac{\partial S_0}{\partial \varepsilon_F} \dfrac{\partial \delta\varepsilon_F (x)}{\partial x}+
\dfrac{\partial S_0}{\partial T} \dfrac{\partial \delta T(x)}{\partial x}\:,
\end{equation} where $\sigma_0$ and $S_0$ are values obtained from Eq.~\eqref{conducivity_diff}
at thermal equilibrium for $V(x)=0$.

Equation~\eqref{zero_current} gives us an expression for the correction to the Fermi energy

\begin{equation}
\delta \varepsilon_F=
-\dfrac{\partial S_0/\partial T}{\sigma_0}\delta T+\delta \tilde{\varepsilon}\:.
\end{equation}
Here, $\delta \tilde{\varepsilon}$ is a constant independent of $x$,
which takes into account that the radiation does not change the average value of the carrier density.
Spatial temperature oscillations can be found using the energy balance

 \[
 |E(x)|^2 \dfrac{\sigma_0}{1+\omega^2 \tau_p^2}=\dfrac{\delta T(x)}{\tau_T}\:,
 \]
 where $\tau_T$ is the temperature relaxation time.
Finally using the relation for the equilibrium distribution function
${\partial f^{(0)}_{\bm{p},\nu }/\partial T=(\partial f^{(0)}_{\bm{p},\nu }/\partial \varepsilon_{\bm{p},\nu})(\varepsilon_{\bm{p},\nu}-\varepsilon_F)/T } $
we obtain the Seebeck ratchet current

\begin{equation}
\label{final_current}
j_x=-\dfrac{1}{e} \dfrac{\sigma_0 \tau_T}{1+\omega^2 \tau_p^2} \left(\dfrac{\partial \sigma_0}{\partial T}-\dfrac{\pi^2}{6}T\left(\dfrac{\partial \sigma_0}{\partial \varepsilon_F}\right)^2\dfrac{1}{\sigma_0}\right) \overline{|E(x)|^2 \dfrac{\partial V}{\partial x}}\:.
\end{equation}

\begin{figure}[h]   %FIG 6

\begin{center}
\includegraphics[width=1.0\linewidth]{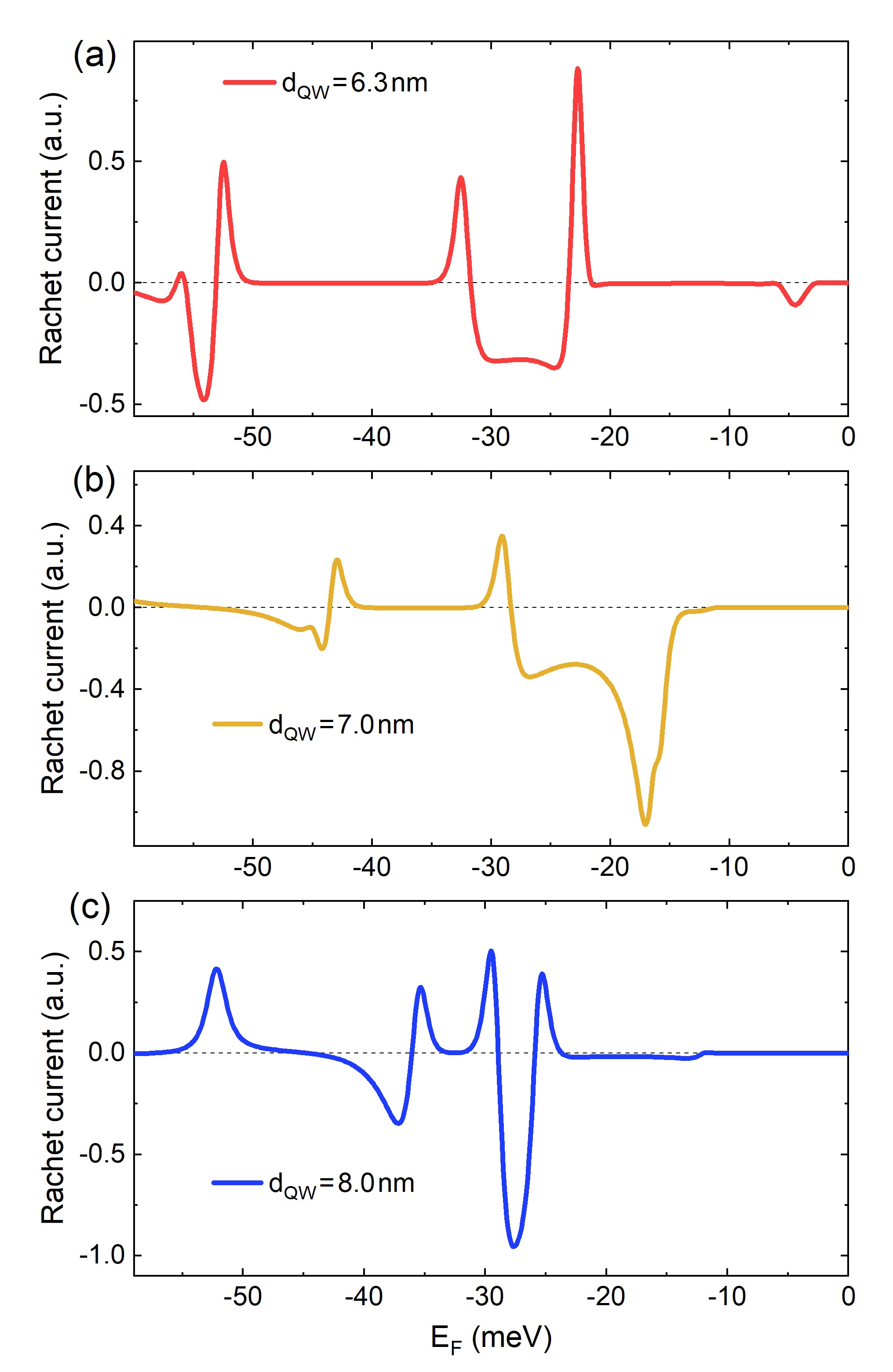}
\end{center}
\caption{Dependencies of the Seebeck ratchet current on the Fermi energy for HgTe quantum wells of different widths.
Photocurrents are calculated after Eq.~\eqref{final_current} using DoS and electron energy spectra from Fig.~\ref{fig_spectrum}.}
\label{fig_ratchet_current}
\end{figure}

To estimate the photocurrent, we make some simplifications: we assume that the momentum relaxation rate is proportional to the number of states where charge carriers can scatter. Similar assumptions can be made about $\tau_T$, so that both $\tau_T^{-1}$, $\tau_p^{-1}$ $\propto g(\varepsilon_{\bm{p},\nu})$, where $g(\varepsilon_{p,\nu})$ is the density of states. We also ignore the anisotropy of the charge carriers spectrum in HgTe quantum wells and take into account only its isotropic part. Figure~\ref{fig_ratchet_current} shows the dependencies of the ratchet current for different quantum wells at $T=4.2$~K. It is clearly seen that the ratchet current is highly sensitive to the details of the energy spectrum. A complex structure is observed in the dependence of the ratchet current on the Fermi energy, which follows the DoS. When the Fermi energy is close to the abrupt change in the density of states or one of the Van Hove singularities, the Seebeck ratchet current is drastically increased or changes its direction. The latter is caused by the large change of the derivative of  conductivity with respect to Fermi energy and temperature for these energies.

The developed theory well describes the main features of the experimental results. Indeed, the theoretical model shows both i) singularities and ii) sign changes across the voltage/energy range with rough agreement with those observed in the experiments (see Figs.~\ref{Fig:8},~\ref{Fig:6} and~\ref{Fig:8}). Therefore, we want to emphasize that despite several simplifications presented model provides evidence of the main responsible physical phenomena. Quantitative comparisons, however, require several model refinements.  First of all, the electrostatic problem for the structure should be solved to calculate the Fermi energy and potential $V(x)$ dependence on the two gate voltages applied to the DGG. $\tau_p$, $\tau_T$ should be accurately calculated for different wave vectors and band indices both for scattering by impurities and phonons taking into account spectrum anisotropy and  that wave function of carries in QW has contributions in the $\Gamma_7$, $\Gamma_8$, and $\Gamma_6$ bands each with its own envelope functions along the $z$-axis. The accurate treatment of relaxation times will significantly alter the ratchet current dependencies shown in Fig.~\ref{fig_ratchet_current}. Lastly, when the Fermi energy exactly matches the energy of one of the Van Hove singularities, the theory loses its applicability, since the difference between electron kinetic energy and local extreme in the energy spectrum goes to zero, and $V(x)$ can't be considered small. As a result, the ratchet current should be expressed in all orders in $V(x)$.  These more precise results can be obtained numerically - but they will not bring new /deeper physical understanding.

\section{Conclusions}

To summarize, different kinds of THz ratchet effects, including the circular, linear, and polarization independent ratchets,  are observed in HgTe-based DGG structures. In comparison with common semiconductors,  HgTe/HgCdTe QW heterostructures exhibit unconventional band structure properties, correlated to quantum well thickness. Measurements on devices with different QW widths demonstrated that the magnitude of the circular ratchet effect  increaseswhen reducing   the QW width and is maximal in structures with critical QW thickness where  the energy gap is close to zero. A further drastic increase of the  ratchet current magnitude is obtained by applying high negative top gate voltages resulting in $p-$type conductivity beneath the gates. The enhancement of the photoresponse as well as the observed sign-alternating oscillations with   gate voltage are shown to be caused by the complex valence band structure of HgTe-based QWs. Our study of ratchet currents in HgTe DGG devices paves the way to develop THz detectors with substantial increased responsivity and opens up  an access to the analysis of energy dispersion details, including the properties of the valence band.

 \section*{Acknowledgments} 
 
 The authors thank L.E.~Golub and M.V.~Durnev for valuable discussions. The support of  the DFG-RFBR project (Ga501/18, RFBR  project 21-52-12015), the Elite Network of Bavaria (K-NW-2013-247), and the Volkswagen Stiftung Program (97738) is gratefully acknowledged. I.Y., W.K. and S.D.G. thank the  IRAP  Programme  of  the Foundation   for   Polish Science   (grant   MAB/2018/9, project CENTERA). A.K. and T.W. thank the IRAP Programme of the Foundation for Polish Science (grant MAB/2017/1, project MagTop). G.V.B. acknowledges the support from the ''BASIS'' foundation. D.W. acknowledges funding by the European Research Council under the European Union's Horizon 2020 research and innovation program (Grant Agreement No. 787515, 253 Pro-Motion).

\bibliography{references}

\begin{thebibliography}{42}%
\makeatletter
\providecommand \@ifxundefined [1]{%
 \@ifx{#1\undefined}
}%
\providecommand \@ifnum [1]{%
 \ifnum #1\expandafter \@firstoftwo
 \else \expandafter \@secondoftwo
 \fi
}%
\providecommand \@ifx [1]{%
 \ifx #1\expandafter \@firstoftwo
 \else \expandafter \@secondoftwo
 \fi
}%
\providecommand \natexlab [1]{#1}%
\providecommand \enquote  [1]{``#1''}%
\providecommand \bibnamefont  [1]{#1}%
\providecommand \bibfnamefont [1]{#1}%
\providecommand \citenamefont [1]{#1}%
\providecommand \href@noop [0]{\@secondoftwo}%
\providecommand \href [0]{\begingroup \@sanitize@url \@href}%
\providecommand \@href[1]{\@@startlink{#1}\@@href}%
\providecommand \@@href[1]{\endgroup#1\@@endlink}%
\providecommand \@sanitize@url [0]{\catcode `\\12\catcode `\$12\catcode
  `\&12\catcode `\#12\catcode `\^12\catcode `\_12\catcode `\%12\relax}%
\providecommand \@@startlink[1]{}%
\providecommand \@@endlink[0]{}%
\providecommand \url  [0]{\begingroup\@sanitize@url \@url }%
\providecommand \@url [1]{\endgroup\@href {#1}{\urlprefix }}%
\providecommand \urlprefix  [0]{URL }%
\providecommand \Eprint [0]{\href }%
\providecommand \doibase [0]{https://doi.org/}%
\providecommand \selectlanguage [0]{\@gobble}%
\providecommand \bibinfo  [0]{\@secondoftwo}%
\providecommand \bibfield  [0]{\@secondoftwo}%
\providecommand \translation [1]{[#1]}%
\providecommand \BibitemOpen [0]{}%
\providecommand \bibitemStop [0]{}%
\providecommand \bibitemNoStop [0]{.\EOS\space}%
\providecommand \EOS [0]{\spacefactor3000\relax}%
\providecommand \BibitemShut  [1]{\csname bibitem#1\endcsname}%
\let\auto@bib@innerbib\@empty
%</preamble>
\bibitem [{\citenamefont {Capper}(1997)}]{Capper1997}%
  \BibitemOpen
  \bibfield  {author} {\bibinfo {author} {\bibfnamefont {P.}~\bibnamefont
  {Capper}},\ }\href
  {https://www.ebook.de/de/product/6732149/narrow_gap_ii_vi_compounds_for_optoelectronic_and_electromagnetic_applications.html}
  {\emph {\bibinfo {title} {Narrow-gap II-VI Compounds for Optoelectronic and
  Electromagnetic Applications}}}\ (\bibinfo  {publisher} {Springer US},\
  \bibinfo {address} {New York City},\ \bibinfo {year} {1997})\BibitemShut
  {NoStop}%
\bibitem [{\citenamefont {Norton}(2002)}]{Norton2002}%
  \BibitemOpen
  \bibfield  {author} {\bibinfo {author} {\bibfnamefont {P.}~\bibnamefont
  {Norton}},\ }\href@noop {} {\bibfield  {journal} {\bibinfo  {journal}
  {Opto-Electron. Rev.}\ }\textbf {\bibinfo {volume} {10}},\ \bibinfo {pages}
  {159} (\bibinfo {year} {2002})}\BibitemShut {NoStop}%
\bibitem [{\citenamefont {Henini}\ and\ \citenamefont
  {Razeghi}(2002)}]{Henini2002}%
  \BibitemOpen
  \bibfield  {author} {\bibinfo {author} {\bibfnamefont {M.}~\bibnamefont
  {Henini}}\ and\ \bibinfo {author} {\bibfnamefont {M.}~\bibnamefont
  {Razeghi}},\ }\href {https://doi.org/10.1016/b978-1-85617-388-9.x5000-x}
  {\emph {\bibinfo {title} {Handbook of Infra-red Detection Technologies}}}\
  (\bibinfo  {publisher} {Elsevier Science},\ \bibinfo {address} {Amsterdam},\
  \bibinfo {year} {2002})\BibitemShut {NoStop}%
\bibitem [{\citenamefont {Rogalski}(2005)}]{Rogalski2005}%
  \BibitemOpen
  \bibfield  {author} {\bibinfo {author} {\bibfnamefont {A.}~\bibnamefont
  {Rogalski}},\ }\href {https://doi.org/10.1088/0034-4885/68/10/r01} {\bibfield
   {journal} {\bibinfo  {journal} {Rep. Prog. Phys.}\ }\textbf {\bibinfo
  {volume} {68}},\ \bibinfo {pages} {2267} (\bibinfo {year}
  {2005})}\BibitemShut {NoStop}%
\bibitem [{\citenamefont {Dvoretsky}\ \emph {et~al.}(2010)\citenamefont
  {Dvoretsky}, \citenamefont {Mikhailov}, \citenamefont {Sidorov},
  \citenamefont {Shvets}, \citenamefont {Danilov}, \citenamefont {Wittman},\
  and\ \citenamefont {Ganichev}}]{Dvoretsky2010}%
  \BibitemOpen
  \bibfield  {author} {\bibinfo {author} {\bibfnamefont {S.}~\bibnamefont
  {Dvoretsky}}, \bibinfo {author} {\bibfnamefont {N.}~\bibnamefont
  {Mikhailov}}, \bibinfo {author} {\bibfnamefont {Y.}~\bibnamefont {Sidorov}},
  \bibinfo {author} {\bibfnamefont {V.}~\bibnamefont {Shvets}}, \bibinfo
  {author} {\bibfnamefont {S.}~\bibnamefont {Danilov}}, \bibinfo {author}
  {\bibfnamefont {B.}~\bibnamefont {Wittman}},\ and\ \bibinfo {author}
  {\bibfnamefont {S.}~\bibnamefont {Ganichev}},\ }\href@noop {} {\bibfield
  {journal} {\bibinfo  {journal} {Journal of Electronic Materials}\ }\textbf
  {\bibinfo {volume} {39}},\ \bibinfo {pages} {918} (\bibinfo {year}
  {2010})}\BibitemShut {NoStop}%
\bibitem [{\citenamefont {Downs}\ and\ \citenamefont
  {Vandervelde}(2013)}]{Downs2013}%
  \BibitemOpen
  \bibfield  {author} {\bibinfo {author} {\bibfnamefont {C.}~\bibnamefont
  {Downs}}\ and\ \bibinfo {author} {\bibfnamefont {T.~E.}\ \bibnamefont
  {Vandervelde}},\ }\href@noop {} {\bibfield  {journal} {\bibinfo  {journal}
  {Sensors}\ }\textbf {\bibinfo {volume} {13}},\ \bibinfo {pages} {5054}
  (\bibinfo {year} {2013})}\BibitemShut {NoStop}%
\bibitem [{\citenamefont {Rogalski}(2018)}]{Rogalski2018}%
  \BibitemOpen
  \bibfield  {author} {\bibinfo {author} {\bibfnamefont {A.}~\bibnamefont
  {Rogalski}},\ }\href
  {https://www.ebook.de/de/product/35182544/antoni_rogalski_infrared_and_terahertz_detectors_third_edition.html}
  {\emph {\bibinfo {title} {Infrared and Terahertz Detectors, Third Edition}}}\
  (\bibinfo  {publisher} {Taylor \& Francis Ltd},\ \bibinfo {address}
  {Abingdon},\ \bibinfo {year} {2018})\BibitemShut {NoStop}%
\bibitem [{\citenamefont {Vanamala}\ \emph {et~al.}(2019)\citenamefont
  {Vanamala}, \citenamefont {Santiago},\ and\ \citenamefont
  {Das}}]{Vanamala2019}%
  \BibitemOpen
  \bibfield  {author} {\bibinfo {author} {\bibfnamefont {N.}~\bibnamefont
  {Vanamala}}, \bibinfo {author} {\bibfnamefont {K.~C.}\ \bibnamefont
  {Santiago}},\ and\ \bibinfo {author} {\bibfnamefont {N.~C.}\ \bibnamefont
  {Das}},\ }\href@noop {} {\bibfield  {journal} {\bibinfo  {journal} {AIP
  Advances}\ }\textbf {\bibinfo {volume} {9}},\ \bibinfo {pages} {025113}
  (\bibinfo {year} {2019})}\BibitemShut {NoStop}%
\bibitem [{\citenamefont {Moore}\ and\ \citenamefont
  {Orenstein}(2010)}]{Moore2010}%
  \BibitemOpen
  \bibfield  {author} {\bibinfo {author} {\bibfnamefont {J.~E.}\ \bibnamefont
  {Moore}}\ and\ \bibinfo {author} {\bibfnamefont {J.}~\bibnamefont
  {Orenstein}},\ }\href {https://doi.org/10.1103/physrevlett.105.026805}
  {\bibfield  {journal} {\bibinfo  {journal} {Phys. Rev. Lett.}\ }\textbf
  {\bibinfo {volume} {105}},\ \bibinfo {pages} {026805} (\bibinfo {year}
  {2010})}\BibitemShut {NoStop}%
\bibitem [{\citenamefont {Hasan}\ and\ \citenamefont {Kane}(2010)}]{Hasan2010}%
  \BibitemOpen
  \bibfield  {author} {\bibinfo {author} {\bibfnamefont {M.~Z.}\ \bibnamefont
  {Hasan}}\ and\ \bibinfo {author} {\bibfnamefont {C.~L.}\ \bibnamefont
  {Kane}},\ }\href {https://doi.org/10.1103/RevModPhys.82.3045} {\bibfield
  {journal} {\bibinfo  {journal} {Rev. Mod. Phys.}\ }\textbf {\bibinfo {volume}
  {82}},\ \bibinfo {pages} {3045} (\bibinfo {year} {2010})}\BibitemShut
  {NoStop}%
\bibitem [{\citenamefont {Zhang}\ \emph {et~al.}(2011)\citenamefont {Zhang},
  \citenamefont {Chang}, \citenamefont {Zhang}, \citenamefont {Wen},
  \citenamefont {Feng}, \citenamefont {Li}, \citenamefont {Liu}, \citenamefont
  {He}, \citenamefont {Wang}, \citenamefont {Chen}, \citenamefont {Xue},
  \citenamefont {Ma},\ and\ \citenamefont {Wang}}]{Zhang2011}%
  \BibitemOpen
  \bibfield  {author} {\bibinfo {author} {\bibfnamefont {J.}~\bibnamefont
  {Zhang}}, \bibinfo {author} {\bibfnamefont {C.-Z.}\ \bibnamefont {Chang}},
  \bibinfo {author} {\bibfnamefont {Z.}~\bibnamefont {Zhang}}, \bibinfo
  {author} {\bibfnamefont {J.}~\bibnamefont {Wen}}, \bibinfo {author}
  {\bibfnamefont {X.}~\bibnamefont {Feng}}, \bibinfo {author} {\bibfnamefont
  {K.}~\bibnamefont {Li}}, \bibinfo {author} {\bibfnamefont {M.}~\bibnamefont
  {Liu}}, \bibinfo {author} {\bibfnamefont {K.}~\bibnamefont {He}}, \bibinfo
  {author} {\bibfnamefont {L.}~\bibnamefont {Wang}}, \bibinfo {author}
  {\bibfnamefont {X.}~\bibnamefont {Chen}}, \bibinfo {author} {\bibfnamefont
  {Q.-K.}\ \bibnamefont {Xue}}, \bibinfo {author} {\bibfnamefont
  {X.}~\bibnamefont {Ma}},\ and\ \bibinfo {author} {\bibfnamefont
  {Y.}~\bibnamefont {Wang}},\ }\href {https://doi.org/10.1038/ncomms1588}
  {\bibfield  {journal} {\bibinfo  {journal} {Nat. Commun.}\ }\textbf {\bibinfo
  {volume} {2}},\ \bibinfo {pages} {574} (\bibinfo {year} {2011})}\BibitemShut
  {NoStop}%
\bibitem [{\citenamefont {Becker}\ \emph {et~al.}(2003)\citenamefont {Becker},
  \citenamefont {Zhang}, \citenamefont {Pfeuffer-Jeschke}, \citenamefont
  {Ortner}, \citenamefont {Hock}, \citenamefont {Landwehr},\ and\ \citenamefont
  {Molenkamp}}]{Becker2003}%
  \BibitemOpen
  \bibfield  {author} {\bibinfo {author} {\bibfnamefont {C.}~\bibnamefont
  {Becker}}, \bibinfo {author} {\bibfnamefont {X.}~\bibnamefont {Zhang}},
  \bibinfo {author} {\bibfnamefont {A.}~\bibnamefont {Pfeuffer-Jeschke}},
  \bibinfo {author} {\bibfnamefont {K.}~\bibnamefont {Ortner}}, \bibinfo
  {author} {\bibfnamefont {V.}~\bibnamefont {Hock}}, \bibinfo {author}
  {\bibfnamefont {G.}~\bibnamefont {Landwehr}},\ and\ \bibinfo {author}
  {\bibfnamefont {L.}~\bibnamefont {Molenkamp}},\ }\href@noop {} {\bibfield
  {journal} {\bibinfo  {journal} {Journal of superconductivity}\ }\textbf
  {\bibinfo {volume} {16}},\ \bibinfo {pages} {625} (\bibinfo {year}
  {2003})}\BibitemShut {NoStop}%
\bibitem [{\citenamefont {Konig}\ \emph {et~al.}(2007)\citenamefont {Konig},
  \citenamefont {Wiedmann}, \citenamefont {Brune}, \citenamefont {Roth},
  \citenamefont {Buhmann}, \citenamefont {Molenkamp}, \citenamefont {Qi},\ and\
  \citenamefont {Zhang}}]{Koenig2007}%
  \BibitemOpen
  \bibfield  {author} {\bibinfo {author} {\bibfnamefont {M.}~\bibnamefont
  {Konig}}, \bibinfo {author} {\bibfnamefont {S.}~\bibnamefont {Wiedmann}},
  \bibinfo {author} {\bibfnamefont {C.}~\bibnamefont {Brune}}, \bibinfo
  {author} {\bibfnamefont {A.}~\bibnamefont {Roth}}, \bibinfo {author}
  {\bibfnamefont {H.}~\bibnamefont {Buhmann}}, \bibinfo {author} {\bibfnamefont
  {L.~W.}\ \bibnamefont {Molenkamp}}, \bibinfo {author} {\bibfnamefont {X.-L.}\
  \bibnamefont {Qi}},\ and\ \bibinfo {author} {\bibfnamefont {S.-C.}\
  \bibnamefont {Zhang}},\ }\href@noop {} {\bibfield  {journal} {\bibinfo
  {journal} {Science}\ }\textbf {\bibinfo {volume} {318}},\ \bibinfo {pages}
  {766} (\bibinfo {year} {2007})}\BibitemShut {NoStop}%
\bibitem [{\citenamefont {Roth}\ \emph {et~al.}(2009)\citenamefont {Roth},
  \citenamefont {Br{\"u}ne}, \citenamefont {Buhmann}, \citenamefont
  {Molenkamp}, \citenamefont {Maciejko}, \citenamefont {Qi},\ and\
  \citenamefont {Zhang}}]{Roth2009}%
  \BibitemOpen
  \bibfield  {author} {\bibinfo {author} {\bibfnamefont {A.}~\bibnamefont
  {Roth}}, \bibinfo {author} {\bibfnamefont {C.}~\bibnamefont {Br{\"u}ne}},
  \bibinfo {author} {\bibfnamefont {H.}~\bibnamefont {Buhmann}}, \bibinfo
  {author} {\bibfnamefont {L.~W.}\ \bibnamefont {Molenkamp}}, \bibinfo {author}
  {\bibfnamefont {J.}~\bibnamefont {Maciejko}}, \bibinfo {author}
  {\bibfnamefont {X.-L.}\ \bibnamefont {Qi}},\ and\ \bibinfo {author}
  {\bibfnamefont {S.-C.}\ \bibnamefont {Zhang}},\ }\href@noop {} {\bibfield
  {journal} {\bibinfo  {journal} {Science}\ }\textbf {\bibinfo {volume}
  {325}},\ \bibinfo {pages} {294} (\bibinfo {year} {2009})}\BibitemShut
  {NoStop}%
\bibitem [{\citenamefont {Dantscher}\ \emph {et~al.}(2015)\citenamefont
  {Dantscher}, \citenamefont {Kozlov}, \citenamefont {Olbrich}, \citenamefont
  {Zoth}, \citenamefont {Faltermeier}, \citenamefont {Lindner}, \citenamefont
  {Budkin}, \citenamefont {Tarasenko}, \citenamefont {Bel'kov}, \citenamefont
  {Kvon}, \citenamefont {Mikhailov}, \citenamefont {Dvoretsky}, \citenamefont
  {Weiss}, \citenamefont {Jenichen},\ and\ \citenamefont
  {Ganichev}}]{Dantscher2015}%
  \BibitemOpen
  \bibfield  {author} {\bibinfo {author} {\bibfnamefont {K.-M.}\ \bibnamefont
  {Dantscher}}, \bibinfo {author} {\bibfnamefont {D.~A.}\ \bibnamefont
  {Kozlov}}, \bibinfo {author} {\bibfnamefont {P.}~\bibnamefont {Olbrich}},
  \bibinfo {author} {\bibfnamefont {C.}~\bibnamefont {Zoth}}, \bibinfo {author}
  {\bibfnamefont {P.}~\bibnamefont {Faltermeier}}, \bibinfo {author}
  {\bibfnamefont {M.}~\bibnamefont {Lindner}}, \bibinfo {author} {\bibfnamefont
  {G.~V.}\ \bibnamefont {Budkin}}, \bibinfo {author} {\bibfnamefont {S.~A.}\
  \bibnamefont {Tarasenko}}, \bibinfo {author} {\bibfnamefont {V.~V.}\
  \bibnamefont {Bel'kov}}, \bibinfo {author} {\bibfnamefont {Z.~D.}\
  \bibnamefont {Kvon}}, \bibinfo {author} {\bibfnamefont {N.~N.}\ \bibnamefont
  {Mikhailov}}, \bibinfo {author} {\bibfnamefont {S.~A.}\ \bibnamefont
  {Dvoretsky}}, \bibinfo {author} {\bibfnamefont {D.}~\bibnamefont {Weiss}},
  \bibinfo {author} {\bibfnamefont {B.}~\bibnamefont {Jenichen}},\ and\
  \bibinfo {author} {\bibfnamefont {S.~D.}\ \bibnamefont {Ganichev}},\
  }\href@noop {} {\bibfield  {journal} {\bibinfo  {journal} {Phys. Rev. B}\
  }\textbf {\bibinfo {volume} {92}},\ \bibinfo {pages} {165314} (\bibinfo
  {year} {2015})}\BibitemShut {NoStop}%
\bibitem [{\citenamefont {Dantscher}\ \emph {et~al.}(2017)\citenamefont
  {Dantscher}, \citenamefont {Kozlov}, \citenamefont {Scherr}, \citenamefont
  {Gebert}, \citenamefont {B{\"a}renf{\"a}nger}, \citenamefont {Durnev},
  \citenamefont {Tarasenko}, \citenamefont {Bel'kov}, \citenamefont
  {Mikhailov}, \citenamefont {Dvoretsky} \emph {et~al.}}]{Dantscher2017}%
  \BibitemOpen
  \bibfield  {author} {\bibinfo {author} {\bibfnamefont {K.-M.}\ \bibnamefont
  {Dantscher}}, \bibinfo {author} {\bibfnamefont {D.}~\bibnamefont {Kozlov}},
  \bibinfo {author} {\bibfnamefont {M.}~\bibnamefont {Scherr}}, \bibinfo
  {author} {\bibfnamefont {S.}~\bibnamefont {Gebert}}, \bibinfo {author}
  {\bibfnamefont {J.}~\bibnamefont {B{\"a}renf{\"a}nger}}, \bibinfo {author}
  {\bibfnamefont {M.}~\bibnamefont {Durnev}}, \bibinfo {author} {\bibfnamefont
  {S.}~\bibnamefont {Tarasenko}}, \bibinfo {author} {\bibfnamefont
  {V.}~\bibnamefont {Bel'kov}}, \bibinfo {author} {\bibfnamefont
  {N.}~\bibnamefont {Mikhailov}}, \bibinfo {author} {\bibfnamefont
  {S.}~\bibnamefont {Dvoretsky}}, \emph {et~al.},\ }\href
  {https://doi.org/10.1103/PhysRevB.92.165314} {\bibfield  {journal} {\bibinfo
  {journal} {Phys. Rev. B}\ }\textbf {\bibinfo {volume} {95}},\ \bibinfo
  {pages} {201103} (\bibinfo {year} {2017})}\BibitemShut {NoStop}%
\bibitem [{\citenamefont {Wittmann}\ \emph {et~al.}(2010)\citenamefont
  {Wittmann}, \citenamefont {Danilov}, \citenamefont {Bel'kov}, \citenamefont
  {Tarasenko}, \citenamefont {Novik}, \citenamefont {Buhmann}, \citenamefont
  {Br\"une}, \citenamefont {Molenkamp}, \citenamefont {Kvon}, \citenamefont
  {Mikhailov}, \citenamefont {Dvoretsky}, \citenamefont {Vinh}, \citenamefont
  {van~der Meer}, \citenamefont {Murdin},\ and\ \citenamefont
  {Ganichev}}]{Wittmann2010}%
  \BibitemOpen
  \bibfield  {author} {\bibinfo {author} {\bibfnamefont {B.}~\bibnamefont
  {Wittmann}}, \bibinfo {author} {\bibfnamefont {S.~N.}\ \bibnamefont
  {Danilov}}, \bibinfo {author} {\bibfnamefont {V.~V.}\ \bibnamefont
  {Bel'kov}}, \bibinfo {author} {\bibfnamefont {S.~A.}\ \bibnamefont
  {Tarasenko}}, \bibinfo {author} {\bibfnamefont {E.~G.}\ \bibnamefont
  {Novik}}, \bibinfo {author} {\bibfnamefont {H.}~\bibnamefont {Buhmann}},
  \bibinfo {author} {\bibfnamefont {C.}~\bibnamefont {Br\"une}}, \bibinfo
  {author} {\bibfnamefont {L.~W.}\ \bibnamefont {Molenkamp}}, \bibinfo {author}
  {\bibfnamefont {Z.~D.}\ \bibnamefont {Kvon}}, \bibinfo {author}
  {\bibfnamefont {N.~N.}\ \bibnamefont {Mikhailov}}, \bibinfo {author}
  {\bibfnamefont {S.~A.}\ \bibnamefont {Dvoretsky}}, \bibinfo {author}
  {\bibfnamefont {N.~Q.}\ \bibnamefont {Vinh}}, \bibinfo {author}
  {\bibfnamefont {A.~F.~G.}\ \bibnamefont {van~der Meer}}, \bibinfo {author}
  {\bibfnamefont {B.}~\bibnamefont {Murdin}},\ and\ \bibinfo {author}
  {\bibfnamefont {S.~D.}\ \bibnamefont {Ganichev}},\ }\href@noop {} {\bibfield
  {journal} {\bibinfo  {journal} {Semicond. Sci. Technol.}\ }\textbf {\bibinfo
  {volume} {25}},\ \bibinfo {pages} {095005} (\bibinfo {year}
  {2010})}\BibitemShut {NoStop}%
\bibitem [{\citenamefont {Danilov}\ \emph {et~al.}(2009)\citenamefont
  {Danilov}, \citenamefont {Wittmann}, \citenamefont {Olbrich}, \citenamefont
  {Eder}, \citenamefont {Prettl}, \citenamefont {Golub}, \citenamefont
  {Beregulin}, \citenamefont {Kvon}, \citenamefont {Mikhailov}, \citenamefont
  {Dvoretsky} \emph {et~al.}}]{Danilov2009}%
  \BibitemOpen
  \bibfield  {author} {\bibinfo {author} {\bibfnamefont {S.}~\bibnamefont
  {Danilov}}, \bibinfo {author} {\bibfnamefont {B.}~\bibnamefont {Wittmann}},
  \bibinfo {author} {\bibfnamefont {P.}~\bibnamefont {Olbrich}}, \bibinfo
  {author} {\bibfnamefont {W.}~\bibnamefont {Eder}}, \bibinfo {author}
  {\bibfnamefont {W.}~\bibnamefont {Prettl}}, \bibinfo {author} {\bibfnamefont
  {L.}~\bibnamefont {Golub}}, \bibinfo {author} {\bibfnamefont
  {E.}~\bibnamefont {Beregulin}}, \bibinfo {author} {\bibfnamefont
  {Z.}~\bibnamefont {Kvon}}, \bibinfo {author} {\bibfnamefont {N.}~\bibnamefont
  {Mikhailov}}, \bibinfo {author} {\bibfnamefont {S.}~\bibnamefont
  {Dvoretsky}}, \emph {et~al.},\ }\href@noop {} {\bibfield  {journal} {\bibinfo
   {journal} {Journal of Applied Physics}\ }\textbf {\bibinfo {volume} {105}},\
  \bibinfo {pages} {013106} (\bibinfo {year} {2009})}\BibitemShut {NoStop}%
\bibitem [{\citenamefont {Olbrich}\ \emph {et~al.}(2009)\citenamefont
  {Olbrich}, \citenamefont {Allerdings}, \citenamefont {Bel'kov}, \citenamefont
  {Tarasenko}, \citenamefont {Schuh}, \citenamefont {Wegscheider},
  \citenamefont {Korn}, \citenamefont {Sch\"uller}, \citenamefont {Weiss},\
  and\ \citenamefont {Ganichev}}]{Olbrich2009}%
  \BibitemOpen
  \bibfield  {author} {\bibinfo {author} {\bibfnamefont {P.}~\bibnamefont
  {Olbrich}}, \bibinfo {author} {\bibfnamefont {J.}~\bibnamefont {Allerdings}},
  \bibinfo {author} {\bibfnamefont {V.~V.}\ \bibnamefont {Bel'kov}}, \bibinfo
  {author} {\bibfnamefont {S.~A.}\ \bibnamefont {Tarasenko}}, \bibinfo {author}
  {\bibfnamefont {D.}~\bibnamefont {Schuh}}, \bibinfo {author} {\bibfnamefont
  {W.}~\bibnamefont {Wegscheider}}, \bibinfo {author} {\bibfnamefont
  {T.}~\bibnamefont {Korn}}, \bibinfo {author} {\bibfnamefont {C.}~\bibnamefont
  {Sch\"uller}}, \bibinfo {author} {\bibfnamefont {D.}~\bibnamefont {Weiss}},\
  and\ \bibinfo {author} {\bibfnamefont {S.~D.}\ \bibnamefont {Ganichev}},\
  }\href@noop {} {\bibfield  {journal} {\bibinfo  {journal} {Phys. Rev. B}\
  }\textbf {\bibinfo {volume} {79}},\ \bibinfo {pages} {245329} (\bibinfo
  {year} {2009})}\BibitemShut {NoStop}%
\bibitem [{\citenamefont {Olbrich}\ \emph {et~al.}(2011)\citenamefont
  {Olbrich}, \citenamefont {Karch}, \citenamefont {Ivchenko}, \citenamefont
  {Kamann}, \citenamefont {M\"arz}, \citenamefont {Fehrenbacher}, \citenamefont
  {Weiss},\ and\ \citenamefont {Ganichev}}]{Olbrich2011}%
  \BibitemOpen
  \bibfield  {author} {\bibinfo {author} {\bibfnamefont {P.}~\bibnamefont
  {Olbrich}}, \bibinfo {author} {\bibfnamefont {J.}~\bibnamefont {Karch}},
  \bibinfo {author} {\bibfnamefont {E.~L.}\ \bibnamefont {Ivchenko}}, \bibinfo
  {author} {\bibfnamefont {J.}~\bibnamefont {Kamann}}, \bibinfo {author}
  {\bibfnamefont {B.}~\bibnamefont {M\"arz}}, \bibinfo {author} {\bibfnamefont
  {M.}~\bibnamefont {Fehrenbacher}}, \bibinfo {author} {\bibfnamefont
  {D.}~\bibnamefont {Weiss}},\ and\ \bibinfo {author} {\bibfnamefont {S.~D.}\
  \bibnamefont {Ganichev}},\ }\href
  {https://doi.org/10.1103/physrevb.83.165320} {\bibfield  {journal} {\bibinfo
  {journal} {Phys. Rev. B}\ }\textbf {\bibinfo {volume} {83}},\ \bibinfo
  {pages} {165320} (\bibinfo {year} {2011})}\BibitemShut {NoStop}%
\bibitem [{\citenamefont {Ivchenko}\ and\ \citenamefont
  {Ganichev}(2011)}]{Ivchenko2011}%
  \BibitemOpen
  \bibfield  {author} {\bibinfo {author} {\bibfnamefont {E.~L.}\ \bibnamefont
  {Ivchenko}}\ and\ \bibinfo {author} {\bibfnamefont {S.~D.}\ \bibnamefont
  {Ganichev}},\ }\href {https://doi.org/10.1134/s002136401111004x} {\bibfield
  {journal} {\bibinfo  {journal} {JETP Lett.}\ }\textbf {\bibinfo {volume}
  {93}},\ \bibinfo {pages} {673} (\bibinfo {year} {2011})},\ \bibinfo {note}
  {[Pisma v ZhETF \textbf{93}, 752 (2011)]}\BibitemShut {NoStop}%
\bibitem [{\citenamefont {Popov}\ \emph {et~al.}(2011)\citenamefont {Popov},
  \citenamefont {Fateev}, \citenamefont {Otsuji}, \citenamefont {Meziani},
  \citenamefont {Coquillat},\ and\ \citenamefont {Knap}}]{Popov2011}%
  \BibitemOpen
  \bibfield  {author} {\bibinfo {author} {\bibfnamefont {V.~V.}\ \bibnamefont
  {Popov}}, \bibinfo {author} {\bibfnamefont {D.~V.}\ \bibnamefont {Fateev}},
  \bibinfo {author} {\bibfnamefont {T.}~\bibnamefont {Otsuji}}, \bibinfo
  {author} {\bibfnamefont {Y.~M.}\ \bibnamefont {Meziani}}, \bibinfo {author}
  {\bibfnamefont {D.}~\bibnamefont {Coquillat}},\ and\ \bibinfo {author}
  {\bibfnamefont {W.}~\bibnamefont {Knap}},\ }\href
  {https://doi.org/10.1063/1.3670321} {\bibfield  {journal} {\bibinfo
  {journal} {Appl. Phys. Lett.}\ }\textbf {\bibinfo {volume} {99}},\ \bibinfo
  {pages} {243504} (\bibinfo {year} {2011})}\BibitemShut {NoStop}%
\bibitem [{\citenamefont {Kannan}\ \emph {et~al.}(2011)\citenamefont {Kannan},
  \citenamefont {Bisotto}, \citenamefont {Portal}, \citenamefont {Murali},\
  and\ \citenamefont {Beck}}]{Kannan2011}%
  \BibitemOpen
  \bibfield  {author} {\bibinfo {author} {\bibfnamefont {E.~S.}\ \bibnamefont
  {Kannan}}, \bibinfo {author} {\bibfnamefont {I.}~\bibnamefont {Bisotto}},
  \bibinfo {author} {\bibfnamefont {J.-C.}\ \bibnamefont {Portal}}, \bibinfo
  {author} {\bibfnamefont {R.}~\bibnamefont {Murali}},\ and\ \bibinfo {author}
  {\bibfnamefont {T.~J.}\ \bibnamefont {Beck}},\ }\href
  {https://doi.org/10.1063/1.3590255} {\bibfield  {journal} {\bibinfo
  {journal} {Appl. Phys. Lett.}\ }\textbf {\bibinfo {volume} {98}},\ \bibinfo
  {pages} {193505} (\bibinfo {year} {2011})}\BibitemShut {NoStop}%
\bibitem [{\citenamefont {Otsuji}\ \emph {et~al.}(2013)\citenamefont {Otsuji},
  \citenamefont {Watanabe}, \citenamefont {Tombet}, \citenamefont {Satou},
  \citenamefont {Knap}, \citenamefont {Popov}, \citenamefont {Ryzhii},\ and\
  \citenamefont {Ryzhii}}]{Otsuji2013}%
  \BibitemOpen
  \bibfield  {author} {\bibinfo {author} {\bibfnamefont {T.}~\bibnamefont
  {Otsuji}}, \bibinfo {author} {\bibfnamefont {T.}~\bibnamefont {Watanabe}},
  \bibinfo {author} {\bibfnamefont {S.~A.~B.}\ \bibnamefont {Tombet}}, \bibinfo
  {author} {\bibfnamefont {A.}~\bibnamefont {Satou}}, \bibinfo {author}
  {\bibfnamefont {W.~M.}\ \bibnamefont {Knap}}, \bibinfo {author}
  {\bibfnamefont {V.~V.}\ \bibnamefont {Popov}}, \bibinfo {author}
  {\bibfnamefont {M.}~\bibnamefont {Ryzhii}},\ and\ \bibinfo {author}
  {\bibfnamefont {V.}~\bibnamefont {Ryzhii}},\ }\href
  {https://doi.org/10.1109/tthz.2012.2235911} {\bibfield  {journal} {\bibinfo
  {journal} {IEEE Trans. Terahertz Sci. Technol.}\ }\textbf {\bibinfo {volume}
  {3}},\ \bibinfo {pages} {63} (\bibinfo {year} {2013})}\BibitemShut {NoStop}%
\bibitem [{\citenamefont {Boubanga-Tombet}\ \emph {et~al.}(2014)\citenamefont
  {Boubanga-Tombet}, \citenamefont {Tanimoto}, \citenamefont {Satou},
  \citenamefont {Suemitsu}, \citenamefont {Wang}, \citenamefont {Minamide},
  \citenamefont {Ito}, \citenamefont {Fateev}, \citenamefont {Popov},\ and\
  \citenamefont {Otsuji}}]{Boubanga2014}%
  \BibitemOpen
  \bibfield  {author} {\bibinfo {author} {\bibfnamefont {S.}~\bibnamefont
  {Boubanga-Tombet}}, \bibinfo {author} {\bibfnamefont {Y.}~\bibnamefont
  {Tanimoto}}, \bibinfo {author} {\bibfnamefont {A.}~\bibnamefont {Satou}},
  \bibinfo {author} {\bibfnamefont {T.}~\bibnamefont {Suemitsu}}, \bibinfo
  {author} {\bibfnamefont {Y.}~\bibnamefont {Wang}}, \bibinfo {author}
  {\bibfnamefont {H.}~\bibnamefont {Minamide}}, \bibinfo {author}
  {\bibfnamefont {H.}~\bibnamefont {Ito}}, \bibinfo {author} {\bibfnamefont
  {D.}~\bibnamefont {Fateev}}, \bibinfo {author} {\bibfnamefont
  {V.}~\bibnamefont {Popov}},\ and\ \bibinfo {author} {\bibfnamefont
  {T.}~\bibnamefont {Otsuji}},\ }\href@noop {} {\bibfield  {journal} {\bibinfo
  {journal} {Applied Physics Letters}\ }\textbf {\bibinfo {volume} {104}},\
  \bibinfo {pages} {262104} (\bibinfo {year} {2014})}\BibitemShut {NoStop}%
\bibitem [{\citenamefont {Faltermeier}\ \emph {et~al.}(2015)\citenamefont
  {Faltermeier}, \citenamefont {Olbrich}, \citenamefont {Probst}, \citenamefont
  {Schell}, \citenamefont {Watanabe}, \citenamefont {Boubanga-Tombet},
  \citenamefont {Otsuji},\ and\ \citenamefont {Ganichev}}]{Faltermeier2015}%
  \BibitemOpen
  \bibfield  {author} {\bibinfo {author} {\bibfnamefont {P.}~\bibnamefont
  {Faltermeier}}, \bibinfo {author} {\bibfnamefont {P.}~\bibnamefont
  {Olbrich}}, \bibinfo {author} {\bibfnamefont {W.}~\bibnamefont {Probst}},
  \bibinfo {author} {\bibfnamefont {L.}~\bibnamefont {Schell}}, \bibinfo
  {author} {\bibfnamefont {T.}~\bibnamefont {Watanabe}}, \bibinfo {author}
  {\bibfnamefont {S.~A.}\ \bibnamefont {Boubanga-Tombet}}, \bibinfo {author}
  {\bibfnamefont {T.}~\bibnamefont {Otsuji}},\ and\ \bibinfo {author}
  {\bibfnamefont {S.~D.}\ \bibnamefont {Ganichev}},\ }\href
  {https://doi.org/10.1063/1.4928969} {\bibfield  {journal} {\bibinfo
  {journal} {J. Appl. Phys.}\ }\textbf {\bibinfo {volume} {118}},\ \bibinfo
  {pages} {084301} (\bibinfo {year} {2015})}\BibitemShut {NoStop}%
\bibitem [{\citenamefont {Rupper}\ \emph {et~al.}(2018)\citenamefont {Rupper},
  \citenamefont {Rudin},\ and\ \citenamefont {Shur}}]{Rupper2018}%
  \BibitemOpen
  \bibfield  {author} {\bibinfo {author} {\bibfnamefont {G.}~\bibnamefont
  {Rupper}}, \bibinfo {author} {\bibfnamefont {S.}~\bibnamefont {Rudin}},\ and\
  \bibinfo {author} {\bibfnamefont {M.~S.}\ \bibnamefont {Shur}},\ }\href
  {https://doi.org/10.1103/physrevapplied.9.064007} {\bibfield  {journal}
  {\bibinfo  {journal} {Phys. Rev. Appl.}\ }\textbf {\bibinfo {volume} {9}},\
  \bibinfo {pages} {064007} (\bibinfo {year} {2018})}\BibitemShut {NoStop}%
\bibitem [{\citenamefont {Yu}(2018)}]{Yu2018}%
  \BibitemOpen
  \bibfield  {author} {\bibinfo {author} {\bibfnamefont {A.}~\bibnamefont
  {Yu}},\ }\href {https://doi.org/10.1088/1361-6463/aad942} {\bibfield
  {journal} {\bibinfo  {journal} {J. Phys. D: Appl. Phys.}\ }\textbf {\bibinfo
  {volume} {51}},\ \bibinfo {pages} {395103} (\bibinfo {year}
  {2018})}\BibitemShut {NoStop}%
\bibitem [{\citenamefont {Delgado-Notario}\ \emph {et~al.}(2020)\citenamefont
  {Delgado-Notario}, \citenamefont {Cleric{\`{o}}}, \citenamefont {Diez},
  \citenamefont {Vel{\'{a}}zquez-P{\'{e}}rez}, \citenamefont {Taniguchi},
  \citenamefont {Watanabe}, \citenamefont {Otsuji},\ and\ \citenamefont
  {Meziani}}]{Notario2020}%
  \BibitemOpen
  \bibfield  {author} {\bibinfo {author} {\bibfnamefont {J.~A.}\ \bibnamefont
  {Delgado-Notario}}, \bibinfo {author} {\bibfnamefont {V.}~\bibnamefont
  {Cleric{\`{o}}}}, \bibinfo {author} {\bibfnamefont {E.}~\bibnamefont {Diez}},
  \bibinfo {author} {\bibfnamefont {J.~E.}\ \bibnamefont
  {Vel{\'{a}}zquez-P{\'{e}}rez}}, \bibinfo {author} {\bibfnamefont
  {T.}~\bibnamefont {Taniguchi}}, \bibinfo {author} {\bibfnamefont
  {K.}~\bibnamefont {Watanabe}}, \bibinfo {author} {\bibfnamefont
  {T.}~\bibnamefont {Otsuji}},\ and\ \bibinfo {author} {\bibfnamefont {Y.~M.}\
  \bibnamefont {Meziani}},\ }\href {https://doi.org/10.1063/5.0007249}
  {\bibfield  {journal} {\bibinfo  {journal} {APL Photonics}\ }\textbf
  {\bibinfo {volume} {5}},\ \bibinfo {pages} {066102} (\bibinfo {year}
  {2020})}\BibitemShut {NoStop}%
\bibitem [{\citenamefont {Sai}\ \emph {et~al.}(2021)\citenamefont {Sai},
  \citenamefont {Potashin}, \citenamefont {Szo{\l}a}, \citenamefont
  {Yavorskiy}, \citenamefont {Cywi{\'{n}}ski}, \citenamefont {Prystawko},
  \citenamefont {{\L}usakowski}, \citenamefont {Ganichev}, \citenamefont
  {Rumyantsev}, \citenamefont {Knap},\ and\ \citenamefont
  {Kachorovskii}}]{Sai2021}%
  \BibitemOpen
  \bibfield  {author} {\bibinfo {author} {\bibfnamefont {P.}~\bibnamefont
  {Sai}}, \bibinfo {author} {\bibfnamefont {S.~O.}\ \bibnamefont {Potashin}},
  \bibinfo {author} {\bibfnamefont {M.}~\bibnamefont {Szo{\l}a}}, \bibinfo
  {author} {\bibfnamefont {D.}~\bibnamefont {Yavorskiy}}, \bibinfo {author}
  {\bibfnamefont {G.}~\bibnamefont {Cywi{\'{n}}ski}}, \bibinfo {author}
  {\bibfnamefont {P.}~\bibnamefont {Prystawko}}, \bibinfo {author}
  {\bibfnamefont {J.}~\bibnamefont {{\L}usakowski}}, \bibinfo {author}
  {\bibfnamefont {S.~D.}\ \bibnamefont {Ganichev}}, \bibinfo {author}
  {\bibfnamefont {S.}~\bibnamefont {Rumyantsev}}, \bibinfo {author}
  {\bibfnamefont {W.}~\bibnamefont {Knap}},\ and\ \bibinfo {author}
  {\bibfnamefont {V.~Y.}\ \bibnamefont {Kachorovskii}},\ }\href
  {https://doi.org/10.1103/physrevb.104.045301} {\bibfield  {journal} {\bibinfo
   {journal} {Phys. Rev. B}\ }\textbf {\bibinfo {volume} {104}},\ \bibinfo
  {pages} {045301} (\bibinfo {year} {2021})}\BibitemShut {NoStop}%
\bibitem [{\citenamefont {Olbrich}\ \emph {et~al.}(2016)\citenamefont
  {Olbrich}, \citenamefont {Kamann}, \citenamefont {K{\"o}nig}, \citenamefont
  {Munzert}, \citenamefont {Tutsch}, \citenamefont {Eroms}, \citenamefont
  {Weiss}, \citenamefont {Liu}, \citenamefont {Golub}, \citenamefont {Ivchenko}
  \emph {et~al.}}]{Olbrich2016}%
  \BibitemOpen
  \bibfield  {author} {\bibinfo {author} {\bibfnamefont {P.}~\bibnamefont
  {Olbrich}}, \bibinfo {author} {\bibfnamefont {J.}~\bibnamefont {Kamann}},
  \bibinfo {author} {\bibfnamefont {M.}~\bibnamefont {K{\"o}nig}}, \bibinfo
  {author} {\bibfnamefont {J.}~\bibnamefont {Munzert}}, \bibinfo {author}
  {\bibfnamefont {L.}~\bibnamefont {Tutsch}}, \bibinfo {author} {\bibfnamefont
  {J.}~\bibnamefont {Eroms}}, \bibinfo {author} {\bibfnamefont
  {D.}~\bibnamefont {Weiss}}, \bibinfo {author} {\bibfnamefont {M.-H.}\
  \bibnamefont {Liu}}, \bibinfo {author} {\bibfnamefont {L.}~\bibnamefont
  {Golub}}, \bibinfo {author} {\bibfnamefont {E.}~\bibnamefont {Ivchenko}},
  \emph {et~al.},\ }\href@noop {} {\bibfield  {journal} {\bibinfo  {journal}
  {Physical Review B}\ }\textbf {\bibinfo {volume} {93}},\ \bibinfo {pages}
  {075422} (\bibinfo {year} {2016})}\BibitemShut {NoStop}%
\bibitem [{\citenamefont {Majewicz}\ \emph {et~al.}(2014)\citenamefont
  {Majewicz}, \citenamefont {{\'S}nie{\.z}ek}, \citenamefont {Wojciechowski},
  \citenamefont {Baran}, \citenamefont {Nowicki}, \citenamefont {Wojtowicz},\
  and\ \citenamefont {Wr{\'o}bel}}]{Majewicz2014}%
  \BibitemOpen
  \bibfield  {author} {\bibinfo {author} {\bibfnamefont {M.}~\bibnamefont
  {Majewicz}}, \bibinfo {author} {\bibfnamefont {D.}~\bibnamefont
  {{\'S}nie{\.z}ek}}, \bibinfo {author} {\bibfnamefont {T.}~\bibnamefont
  {Wojciechowski}}, \bibinfo {author} {\bibfnamefont {E.}~\bibnamefont
  {Baran}}, \bibinfo {author} {\bibfnamefont {P.}~\bibnamefont {Nowicki}},
  \bibinfo {author} {\bibfnamefont {T.}~\bibnamefont {Wojtowicz}},\ and\
  \bibinfo {author} {\bibfnamefont {J.}~\bibnamefont {Wr{\'o}bel}},\
  }\href@noop {} {\bibfield  {journal} {\bibinfo  {journal} {Acta Physica
  Polonica, A.}\ }\textbf {\bibinfo {volume} {126}} (\bibinfo {year}
  {2014})}\BibitemShut {NoStop}%
\bibitem [{\citenamefont {Majewicz}(2019)}]{Majewicz2019}%
  \BibitemOpen
  \bibfield  {author} {\bibinfo {author} {\bibfnamefont {M.~M.}\ \bibnamefont
  {Majewicz}},\ }\href@noop {} {\emph {\bibinfo {title} {Wytwarzanie
  nanostruktur i badanie zjawisk transportu w dwuwymiarowych izolatorach
  topologicznych}}}\ (\bibinfo  {publisher} {PhD thesis, Institute of Physics
  PAS},\ \bibinfo {address} {Warsaw},\ \bibinfo {year} {2019})\BibitemShut
  {NoStop}%
\bibitem [{\citenamefont {Shalygin}\ \emph {et~al.}(2007)\citenamefont
  {Shalygin}, \citenamefont {Diehl}, \citenamefont {Hoffmann}, \citenamefont
  {Danilov}, \citenamefont {Herrle}, \citenamefont {Tarasenko}, \citenamefont
  {Schuh}, \citenamefont {Gerl}, \citenamefont {Wegscheider}, \citenamefont
  {Prettl} \emph {et~al.}}]{Shalygin2007}%
  \BibitemOpen
  \bibfield  {author} {\bibinfo {author} {\bibfnamefont {V.}~\bibnamefont
  {Shalygin}}, \bibinfo {author} {\bibfnamefont {H.}~\bibnamefont {Diehl}},
  \bibinfo {author} {\bibfnamefont {C.}~\bibnamefont {Hoffmann}}, \bibinfo
  {author} {\bibfnamefont {S.}~\bibnamefont {Danilov}}, \bibinfo {author}
  {\bibfnamefont {T.}~\bibnamefont {Herrle}}, \bibinfo {author} {\bibfnamefont
  {S.}~\bibnamefont {Tarasenko}}, \bibinfo {author} {\bibfnamefont
  {D.}~\bibnamefont {Schuh}}, \bibinfo {author} {\bibfnamefont
  {C.}~\bibnamefont {Gerl}}, \bibinfo {author} {\bibfnamefont {W.}~\bibnamefont
  {Wegscheider}}, \bibinfo {author} {\bibfnamefont {W.}~\bibnamefont {Prettl}},
  \emph {et~al.},\ }\href@noop {} {\bibfield  {journal} {\bibinfo  {journal}
  {JETP letters}\ }\textbf {\bibinfo {volume} {84}},\ \bibinfo {pages} {570}
  (\bibinfo {year} {2007})}\BibitemShut {NoStop}%
\bibitem [{\citenamefont {Plank}\ \emph {et~al.}(2016)\citenamefont {Plank},
  \citenamefont {Golub}, \citenamefont {Bauer}, \citenamefont {Bel'kov},
  \citenamefont {Herrmann}, \citenamefont {Olbrich}, \citenamefont {Eschbach},
  \citenamefont {Plucinski}, \citenamefont {Schneider}, \citenamefont
  {Kampmeier}, \citenamefont {Lanius}, \citenamefont {Mussler}, \citenamefont
  {Gr\"utzmacher},\ and\ \citenamefont {Ganichev}}]{Plank2016}%
  \BibitemOpen
  \bibfield  {author} {\bibinfo {author} {\bibfnamefont {H.}~\bibnamefont
  {Plank}}, \bibinfo {author} {\bibfnamefont {L.~E.}\ \bibnamefont {Golub}},
  \bibinfo {author} {\bibfnamefont {S.}~\bibnamefont {Bauer}}, \bibinfo
  {author} {\bibfnamefont {V.~V.}\ \bibnamefont {Bel'kov}}, \bibinfo {author}
  {\bibfnamefont {T.}~\bibnamefont {Herrmann}}, \bibinfo {author}
  {\bibfnamefont {P.}~\bibnamefont {Olbrich}}, \bibinfo {author} {\bibfnamefont
  {M.}~\bibnamefont {Eschbach}}, \bibinfo {author} {\bibfnamefont
  {L.}~\bibnamefont {Plucinski}}, \bibinfo {author} {\bibfnamefont {C.~M.}\
  \bibnamefont {Schneider}}, \bibinfo {author} {\bibfnamefont {J.}~\bibnamefont
  {Kampmeier}}, \bibinfo {author} {\bibfnamefont {M.}~\bibnamefont {Lanius}},
  \bibinfo {author} {\bibfnamefont {G.}~\bibnamefont {Mussler}}, \bibinfo
  {author} {\bibfnamefont {D.}~\bibnamefont {Gr\"utzmacher}},\ and\ \bibinfo
  {author} {\bibfnamefont {S.~D.}\ \bibnamefont {Ganichev}},\ }\href
  {https://doi.org/10.1103/PhysRevB.93.125434} {\bibfield  {journal} {\bibinfo
  {journal} {Phys. Rev. B}\ }\textbf {\bibinfo {volume} {93}},\ \bibinfo
  {pages} {125434} (\bibinfo {year} {2016})}\BibitemShut {NoStop}%
\bibitem [{\citenamefont {Bel'kov}\ \emph {et~al.}(2005)\citenamefont
  {Bel'kov}, \citenamefont {Ganichev}, \citenamefont {Ivchenko}, \citenamefont
  {Tarasenko}, \citenamefont {Weber}, \citenamefont {Giglberger}, \citenamefont
  {Olteanu}, \citenamefont {Tranitz}, \citenamefont {Danilov}, \citenamefont
  {Schneider}, \citenamefont {Wegscheider}, \citenamefont {Weiss},\ and\
  \citenamefont {Prettl}}]{Belkov2005}%
  \BibitemOpen
  \bibfield  {author} {\bibinfo {author} {\bibfnamefont {V.~V.}\ \bibnamefont
  {Bel'kov}}, \bibinfo {author} {\bibfnamefont {S.~D.}\ \bibnamefont
  {Ganichev}}, \bibinfo {author} {\bibfnamefont {E.~L.}\ \bibnamefont
  {Ivchenko}}, \bibinfo {author} {\bibfnamefont {S.~A.}\ \bibnamefont
  {Tarasenko}}, \bibinfo {author} {\bibfnamefont {W.}~\bibnamefont {Weber}},
  \bibinfo {author} {\bibfnamefont {S.}~\bibnamefont {Giglberger}}, \bibinfo
  {author} {\bibfnamefont {M.}~\bibnamefont {Olteanu}}, \bibinfo {author}
  {\bibfnamefont {H.~P.}\ \bibnamefont {Tranitz}}, \bibinfo {author}
  {\bibfnamefont {S.~N.}\ \bibnamefont {Danilov}}, \bibinfo {author}
  {\bibfnamefont {P.}~\bibnamefont {Schneider}}, \bibinfo {author}
  {\bibfnamefont {W.}~\bibnamefont {Wegscheider}}, \bibinfo {author}
  {\bibfnamefont {D.}~\bibnamefont {Weiss}},\ and\ \bibinfo {author}
  {\bibfnamefont {W.}~\bibnamefont {Prettl}},\ }\href@noop {} {\bibfield
  {journal} {\bibinfo  {journal} {J. Phys. Condens. Matter}\ }\textbf {\bibinfo
  {volume} {17}},\ \bibinfo {pages} {3405} (\bibinfo {year}
  {2005})}\BibitemShut {NoStop}%
\bibitem [{\citenamefont {Faltermeier}\ \emph {et~al.}(2017)\citenamefont
  {Faltermeier}, \citenamefont {Budkin}, \citenamefont {Unverzagt},
  \citenamefont {Hubmann}, \citenamefont {Pfaller}, \citenamefont {Bel'kov},
  \citenamefont {Golub}, \citenamefont {Ivchenko}, \citenamefont {Adamus},
  \citenamefont {Karczewski}, \citenamefont {Wojtowicz}, \citenamefont {Popov},
  \citenamefont {Fateev}, \citenamefont {Kozlov}, \citenamefont {Weiss},\ and\
  \citenamefont {Ganichev}}]{Faltermeier2017}%
  \BibitemOpen
  \bibfield  {author} {\bibinfo {author} {\bibfnamefont {P.}~\bibnamefont
  {Faltermeier}}, \bibinfo {author} {\bibfnamefont {G.~V.}\ \bibnamefont
  {Budkin}}, \bibinfo {author} {\bibfnamefont {J.}~\bibnamefont {Unverzagt}},
  \bibinfo {author} {\bibfnamefont {S.}~\bibnamefont {Hubmann}}, \bibinfo
  {author} {\bibfnamefont {A.}~\bibnamefont {Pfaller}}, \bibinfo {author}
  {\bibfnamefont {V.~V.}\ \bibnamefont {Bel'kov}}, \bibinfo {author}
  {\bibfnamefont {L.~E.}\ \bibnamefont {Golub}}, \bibinfo {author}
  {\bibfnamefont {E.~L.}\ \bibnamefont {Ivchenko}}, \bibinfo {author}
  {\bibfnamefont {Z.}~\bibnamefont {Adamus}}, \bibinfo {author} {\bibfnamefont
  {G.}~\bibnamefont {Karczewski}}, \bibinfo {author} {\bibfnamefont
  {T.}~\bibnamefont {Wojtowicz}}, \bibinfo {author} {\bibfnamefont {V.~V.}\
  \bibnamefont {Popov}}, \bibinfo {author} {\bibfnamefont {D.~V.}\ \bibnamefont
  {Fateev}}, \bibinfo {author} {\bibfnamefont {D.~A.}\ \bibnamefont {Kozlov}},
  \bibinfo {author} {\bibfnamefont {D.}~\bibnamefont {Weiss}},\ and\ \bibinfo
  {author} {\bibfnamefont {S.~D.}\ \bibnamefont {Ganichev}},\ }\href
  {https://doi.org/10.1103/physrevb.95.155442} {\bibfield  {journal} {\bibinfo
  {journal} {Phys. Rev. B}\ }\textbf {\bibinfo {volume} {95}},\ \bibinfo
  {pages} {155442} (\bibinfo {year} {2017})}\BibitemShut {NoStop}%
\bibitem [{\citenamefont {Budkin}\ \emph {et~al.}(2016)\citenamefont {Budkin},
  \citenamefont {Golub}, \citenamefont {Ivchenko},\ and\ \citenamefont
  {Ganichev}}]{Budkin2016}%
  \BibitemOpen
  \bibfield  {author} {\bibinfo {author} {\bibfnamefont {G.~V.}\ \bibnamefont
  {Budkin}}, \bibinfo {author} {\bibfnamefont {L.~E.}\ \bibnamefont {Golub}},
  \bibinfo {author} {\bibfnamefont {E.~L.}\ \bibnamefont {Ivchenko}},\ and\
  \bibinfo {author} {\bibfnamefont {S.~D.}\ \bibnamefont {Ganichev}},\ }\href
  {https://doi.org/10.1134/s0021364016210074} {\bibfield  {journal} {\bibinfo
  {journal} {JETP Lett.}\ }\textbf {\bibinfo {volume} {104}},\ \bibinfo {pages}
  {649} (\bibinfo {year} {2016})}\BibitemShut {NoStop}%
\bibitem [{\citenamefont {Faltermeier}\ \emph {et~al.}(2018)\citenamefont
  {Faltermeier}, \citenamefont {Budkin}, \citenamefont {Hubmann}, \citenamefont
  {Bel'kov}, \citenamefont {Golub}, \citenamefont {Ivchenko}, \citenamefont
  {Adamus}, \citenamefont {Karczewski}, \citenamefont {Wojtowicz},
  \citenamefont {Kozlov}, \citenamefont {Weiss},\ and\ \citenamefont
  {Ganichev}}]{Faltermeier2018}%
  \BibitemOpen
  \bibfield  {author} {\bibinfo {author} {\bibfnamefont {P.}~\bibnamefont
  {Faltermeier}}, \bibinfo {author} {\bibfnamefont {G.~V.}\ \bibnamefont
  {Budkin}}, \bibinfo {author} {\bibfnamefont {S.}~\bibnamefont {Hubmann}},
  \bibinfo {author} {\bibfnamefont {V.~V.}\ \bibnamefont {Bel'kov}}, \bibinfo
  {author} {\bibfnamefont {L.~E.}\ \bibnamefont {Golub}}, \bibinfo {author}
  {\bibfnamefont {E.~L.}\ \bibnamefont {Ivchenko}}, \bibinfo {author}
  {\bibfnamefont {Z.}~\bibnamefont {Adamus}}, \bibinfo {author} {\bibfnamefont
  {G.}~\bibnamefont {Karczewski}}, \bibinfo {author} {\bibfnamefont
  {T.}~\bibnamefont {Wojtowicz}}, \bibinfo {author} {\bibfnamefont {D.~A.}\
  \bibnamefont {Kozlov}}, \bibinfo {author} {\bibfnamefont {D.}~\bibnamefont
  {Weiss}},\ and\ \bibinfo {author} {\bibfnamefont {S.~D.}\ \bibnamefont
  {Ganichev}},\ }\href {https://doi.org/10.1016/j.physe.2018.04.001} {\bibfield
   {journal} {\bibinfo  {journal} {Physica E}\ }\textbf {\bibinfo {volume}
  {101}},\ \bibinfo {pages} {178} (\bibinfo {year} {2018})}\BibitemShut
  {NoStop}%
\bibitem [{\citenamefont {Hubmann}\ \emph {et~al.}(2020)\citenamefont
  {Hubmann}, \citenamefont {Bel'kov}, \citenamefont {Golub}, \citenamefont
  {Kachorovskii}, \citenamefont {Drienovsky}, \citenamefont {Eroms},
  \citenamefont {Weiss},\ and\ \citenamefont {Ganichev}}]{Hubmann2020}%
  \BibitemOpen
  \bibfield  {author} {\bibinfo {author} {\bibfnamefont {S.}~\bibnamefont
  {Hubmann}}, \bibinfo {author} {\bibfnamefont {V.~V.}\ \bibnamefont
  {Bel'kov}}, \bibinfo {author} {\bibfnamefont {L.~E.}\ \bibnamefont {Golub}},
  \bibinfo {author} {\bibfnamefont {V.~Y.}\ \bibnamefont {Kachorovskii}},
  \bibinfo {author} {\bibfnamefont {M.}~\bibnamefont {Drienovsky}}, \bibinfo
  {author} {\bibfnamefont {J.}~\bibnamefont {Eroms}}, \bibinfo {author}
  {\bibfnamefont {D.}~\bibnamefont {Weiss}},\ and\ \bibinfo {author}
  {\bibfnamefont {S.~D.}\ \bibnamefont {Ganichev}},\ }\href
  {https://doi.org/10.1103/physrevresearch.2.033186} {\bibfield  {journal}
  {\bibinfo  {journal} {Phys. Rev. Research}\ }\textbf {\bibinfo {volume}
  {2}},\ \bibinfo {pages} {033186} (\bibinfo {year} {2020})}\BibitemShut
  {NoStop}%
\bibitem [{\citenamefont {Durnev}(2021)}]{Durnev2021}%
  \BibitemOpen
  \bibfield  {author} {\bibinfo {author} {\bibfnamefont {M.}~\bibnamefont
  {Durnev}},\ }\href@noop {} {\bibfield  {journal} {\bibinfo  {journal} {Phys.
  Rev. B}\ }\textbf {\bibinfo {volume} {104}},\ \bibinfo {pages} {085306}
  (\bibinfo {year} {2021})}\BibitemShut {NoStop}%
\bibitem [{\citenamefont {Nalitov}\ \emph {et~al.}(2012)\citenamefont
  {Nalitov}, \citenamefont {Golub},\ and\ \citenamefont
  {Ivchenko}}]{Nalitov2012}%
  \BibitemOpen
  \bibfield  {author} {\bibinfo {author} {\bibfnamefont {A.}~\bibnamefont
  {Nalitov}}, \bibinfo {author} {\bibfnamefont {L.}~\bibnamefont {Golub}},\
  and\ \bibinfo {author} {\bibfnamefont {E.}~\bibnamefont {Ivchenko}},\
  }\href@noop {} {\bibfield  {journal} {\bibinfo  {journal} {Phys. Rev. B}\
  }\textbf {\bibinfo {volume} {86}},\ \bibinfo {pages} {115301} (\bibinfo
  {year} {2012})}\BibitemShut {NoStop}%
\end{thebibliography}%

\end{document}